\documentclass[printer]{aa}

\newcommand{\asca}{{\it ASCA}}

\newcommand{\ginga}{{\it Ginga}}

\newcommand{\rosat}{{\it ROSAT}}
\newcommand{\xmm}{{\it XMM-Newton}}
\newcommand{\epic}{{\it EPIC}}
\newcommand{\rgs}{{\it RGS}}
\newcommand{\om}{{\it OM}}
\newcommand{\pn}{{\it EPIC-pn}}
\newcommand{\chandra}{{\it Chandra}}

\newcommand{\msol}{$M_{\odot}$}
\newcommand{\msolyr}{$M_{\odot}$\,yr$^{-1}$}
\newcommand{\hi}{H\,{\small{\sc I}}}
\newcommand{\hii}{H\,{\small{\sc II}}}

\newcommand{\halpha}{H$\alpha$}

\usepackage{graphics}
\usepackage{psfig}
\usepackage{subfigure}

\begin{document}

\hyphenation{con-sti-tu-ents there-by mar-gin-al-ly stat-is-tics}

\title{X-ray emission from  NGC\,1808: more than a complex
starburst\thanks{Partially based on observations obtained with XMM-Newton, an ESA science mission with instruments and contributions directly funded by ESA Member States and NASA.}}

\titlerunning{X-ray Emission from  NGC\,1808 }
\authorrunning{E. Jim\'enez-Bail\'on  et al.}

\author{E. Jim\'enez Bail\'on\inst{1}
\and
M. Santos-Lle\'o\inst{1}
\and
M. Dahlem\inst{2}
\and
M. Ehle\inst{1}
\and
J. M. Mas-Hesse\inst{3}
\and
M. Guainazzi\inst{1}
\and
T. M. Heckman\inst{4}
\and
K. A. Weaver\inst{5}
}

\institute{XMM-Newton Science Operations Centre, ESAC, ESA, Apartado 50727,
  E-28080 Madrid, Spain
\and
ATNF/CSIRO, Paul Wild Observatory, Locked Bag 194, Narrabri
  NSW 2390, Australia
\and
Centro de Astrobiolog\'{\i}a (CSIC-INTA), E-28850 Torrej\'on
de Ardoz, Madrid, Spain
\and
Dept. of Physics and Astronomy, Johns Hopkins University,
3400 N. Charles St., Baltimore, MD 21218, USA 
\and
Laboratory for High Energy Astrophysics, NASA's
Goddard Space Flight Center, Greenbelt, MD 20771, USA
}

\offprints{Elena.Jimenez@sciops.esa.int}

\date{Received xx month 200x / Accepted yy month 200y}

\abstract{

Earlier  observations  of   NGC\,1808  in  various  wavebands  (X-ray,
optical, near-infrared,  radio)  provided  evidence for  the
existence of either a starburst or a Seyfert\,2 nucleus. We here present
the results of  multiwavelength  \xmm\ and \chandra\ observations, which
  directly prove the   co-existence  of  thermal  diffuse  plasma   and
non-nuclear unresolved  point-like    sources  associated  with   the
starburst   activity, along with  a Low   Luminosity   Active Galactic  Nucleus
(LLAGN) or an Ultra Luminous X-ray source (ULX).
The  broad bandwidth of  \xmm\ allows  us to  show that the unresolved
nuclear source in NGC\,1808   dominates the hard X-ray spectrum,  while
the  emission in the  soft  regime, below 1   keV,  is dominated by  a
thermal component associated to an extended starburst. Both \epic\ and
\rgs\ data provide reliable detections  of a number of emission  lines
from  heavy elements, with abundances  ranging from roughly 0.7 to 2.2
$Z_\odot$  for different elements.   However, no 6.4 keV Fe  K$\alpha$
fluorescence  line emission was  detected. The analysis of the nuclear
region  of    NGC\,1808  allows  us  to  detect   and  disentangle the
contribution of an unresolved  nuclear X-ray source and  the starburst
region, but  the exact  nature  of the  nucleus remains  unknown.  The
observed         luminosity        of            NGC\,1808          is
$L_{2-10\,keV}=(1.61\pm0.06)\times10^{40}$ erg s$^{-1}$.
A comparison of our \om\ 212\,nm image with a CTIO 4-m telescope 
H$\alpha$ frame shows a good general correspondence between the
 emission from massive stars and warm ionized gas, with minor
deviations near the ends of the bar in NGC\,1808. 
An aditional, very soft thermal spectral component with kT$\simeq$0.1~keV 
has been discovered in the \xmm\ spectral analysis, which most likely 
originates from the halo of NGC\,1808.

\keywords{Galaxies:~active -- Galaxies:~nuclei -- Galaxies:~general --
X-rays --ISM: general -- Galaxies: evolution -- Galaxies: halos 
-- Galaxies: starburst -- individual: NGC1808}
}

\maketitle

\section{Introduction}
\label{par:intro}

NGC\,1808  ($\alpha$(J2000)=05:07:42.34;  $\delta$(J2000)=--37:30:47.0
with an uncertainty of 1\farcs25,  obtained from 2MASS\footnote{  {\it
http://www.ipac.caltech.edu/2mass/}}), classified as an SABb Seyfert~2
galaxy, is located at  a  distance of 10.9   Mpc (H$_0$ =   75  km s\,$^{-1}$ Mpc$^{-1}$,
1\arcsec = 53 pc).
Images in different wavebands    suggest a high star  formation   (SF)
intensity  in the central region  with a diameter of $\sim 20$\arcsec\
(i.e., $\approx  1$   kpc).  There   are  several  optical  hot  spots
associated    with \hii-regions (S\'ersic  \&    Pastoriza 1965).  The
luminous and compact knots detected in radio and infra-red (IR) images
(Saikia et al. 1990, Kotilainen et al. 1996) that do not coincide with
the optical hot   spots  are probably  supernova  remnants (SNRs)  or
complexes of unresolved SNRs. Dust  filaments were found perpendicular
to the  central  disk plane  and   can be explained as   outflowing
material driven by supernovae (SNe; Heckman et al. 1990).
A recent interaction of NGC\,1808 with its companion NGC\,1792 could 
explain both the intense star-formation activity and its peculiar 
morphology (Dahlem et al. 1990, Koribalski et al. 1993).

The   nature of  the  nucleus is  still unclear.   It is classified as
Seyfert~2 based on the   optical  emission lines detected   by
V\'eron-Cetty \& V\'eron   (1985). However, there is also evidence from optical  observations of a hidden
starburst (SB),    based on polarization  measurements  by  Scarrott et al.
(1993).
The nucleus is claimed to have a strong non-stellar component,
because only 10\% of the IR radiation observed by {\it 
ISO} could be interpreted as emission related to star formation
(Siebenmorgen et al. 2001). 
On the  other  hand, soft  X-ray  (0.1--2.4  keV) data  obtained  with
\rosat\ (Dahlem et al.   1994    and Junkes   et al.   1995)   favor an
interpretation in terms of SF-related   emission, but do not  discard
other hypotheses.
Awaki \&  Koyama (1993) interpret their \ginga\ X-ray data 
(1.5--37 keV) as an obscured active galactic nucleus (AGN). However, Awaki et 
al. (1996), using \asca\ (2--10 keV) observations, point out that 
the hard X-ray spectrum could also be the result of starburst 
activity, while the long-term variability from the \ginga\ and 
\asca\ observations again suggests a Seyfert nature of the 
nucleus. 

%

With the advent of the \xmm\ and \chandra\ satellites, the coexistence of 
starburst and AGN   activity  has been detected in   X-rays  in the  nuclei of
several nearby galaxies.   NGC\,4303 presents evidence of the composite
SB/AGN nature  of  the  nuclear  region:  \chandra\ and UV-{\it   HST}
observations  show  that a superstellar cluster  coexists  with a hard
X-ray  source, possibly an  AGN,  in the 3 central   pc of the  galaxy
(Jim\'enez-Bail\'on  et al.  2003, Colina  et al.   2002).  At larger
scales, Persic   et al.   (2004a)  claim  that starburst   and AGN coexist in
NGC\,4666 by   detecting   evidence  of  diffuse   thermal   emission,
high-luminousity X-ray sources, and the presence of an AGN in the X-ray
band. The detailed analysis of our \xmm\ data shown here, combined with
other observations, sheds   new light on  the  nature of  the  nuclear
activity in NGC\,1808 and the associated outflow of gas.

The observations and data reduction are described in 
Sect.~\ref{par:obsred}, our results are presented in
Sect.~\ref{par:results}, followed by our interpretation of the
inner region and circumnuclear emission in Sect.~\ref{par:centre}\ and 
of the more extended X-ray emission outside the nuclear area in 
Sect.~\ref{par:outer}. The conclusions of this work are given in Sect.~\ref{par:summ}.

\section{Observations and data reduction}
\label{par:obsred}

\subsection{\xmm\ data}

Information on the \xmm\ instrumentation is provided by 
Jansen et al. (2001; \xmm\ mission), Str\"uder et al. 
(2001; \pn), Turner et al. (2001; {\it EPIC-MOS}), den 
Herder et al. (2001; \rgs), and Mason et al. (2001; {\it OM}).
Our \xmm\ observation (Obs-Id~0110980801) was performed on April 6, 2002. 
The scheduled exposure time was 40~ks, while effective exposure times 
per instrument are listed below. The {\it EPIC-pn} exposure was in 
extended full frame mode and with the thin filter. 
Both \rgs\ cameras were in spectroscopic mode and the two {\it MOS} cameras in
full frame, with the thin filter. {\it OM} was used in full frame low 
resolution mode with the U, UVW1 (219~nm) and UVW2 (212~nm) broad band 
filters, plus the two (optical and UV) grisms. 

The  data  were  processed  with  the  {\it Science  Analysis
Subsystem}, SAS,~v.5.4.1 (Gabriel et al. 2004) and  the  most up-to-date 
calibration available in February 2003. 
The standard tasks to process data, {\it ep/mchain} and  {\it rgsproc}, 
were run  with default parameters. For the \epic\ instruments, 
checks against pile-up were performed with the {\it epatplot} task. 
No sign of  pile-up was found for any of {\it EPIC-pn}, {\it MOS1}, 
or {\it MOS2} data. Additionally, background flaring time intervals, 
for which the signal-to-noise does not improve by including these 
events, were filtered following the method described in Piconcelli
et al. (2004b). Only single and double events were considered in
the analysis.
For {\it RGS} the  selection was for periods with less than 0.12 
c/s in the background region of the CCD number 9.
The effective exposure times after each filter was applied were 31.4 ks,
39.0 ks,  38.6 ks, and  37.4 ks for {\it EPIC-pn},  the  two {\it MOS}
cameras, and  {\it RGS}, respectively. The  \om\ broad band
data were processed with the SAS {\it omichain} task, which performed all
necessary corrections. Unfortunately,  about  half of the  UVW1 filter
image was  lost due to ground-station  problems, and the UV grism data
was also lost due to telemetry drops.

\subsection{\chandra\ data}

The \chandra\ ACIS ({\it AXAF CCD Imaging Camera}) archival
image, sequence number 700451, was extracted 
from the \chandra\ archive.
The ACIS observation took place on December 19,  2002 with an exposure
time of 43~ks. We  used the science {\it level 2} files, generated
by the pipeline standard processing on December 19, 2002, as stored in
the  archive.  They were  processed with Ciao~3.0.2.  According to the
information available on the
\chandra\ pages, further  corrections should be applied (i.e. reduction
of tap-ringing distortions). However,  for
the purpose of the present work, this was found to be unnecessary.  No
additional corrections were therefore applied.

\subsection{Ground-based H$\alpha$ imagery}

The H$\alpha$ data used here were obtained with the CTIO 4-m
telescope on November 18, 1993. A $2048\times2048$ chip with a 
$0\farcs431$ pixel scale was
used, leading to a total field-of-view of $14\farcm7\times14\farcm7$.

Three exposures of 20 min. each were taken with a redshifted 
H$\alpha$ filter ($\lambda_0$ = 657.8 nm; $\Delta\lambda$ = 0.64 
nm), with a total integration time of 1 hr. 
The continuum subtraction was performed using an $R$-band image 
with a total integration time of 8 min.
The data are extremely sensitive to low surface brightness
emission, but as they are uncalibrated,  will be used only for 
qualitative studies.

\begin{figure*}[htb]
\psfig{file=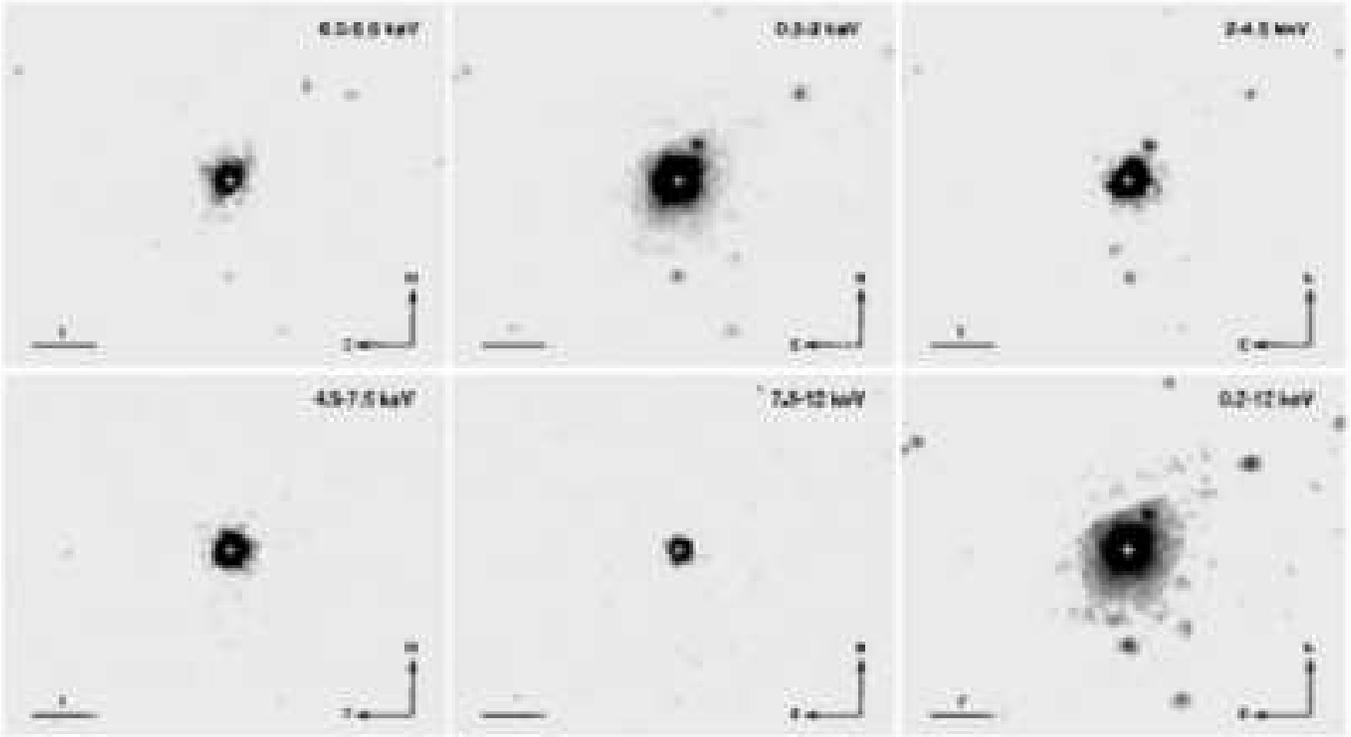,width=180mm,angle=0}
\caption{EPIC images of NGC\,1808 in the following energy
bands (from left to right and from top to bottom): 0.2--0.5 keV, 
0.5--2.0 keV, 2.0--4.5 keV, 4.5--7.5 keV, 7.5--12 keV, and the
total bandpass 0.2--12 keV. All images are displayed on a logarithmic
scale with a threshold of 3$\sigma$ of the mean background value.}
\label{fig:xbands}
\end{figure*}

\section{Results}
\label{par:results}

\subsection{X-ray imaging}

\xmm\ \epic\ X-ray images of NGC\,1808 were produced by grouping 
the photons from  the cleaned event lists  into the sub-bands defined
by the \xmm\ Survey Science Centre (SSC; Watson et  al. 2001) and using
the {\it merge} task in SAS~v.5.4.1, which takes  exposure
time      corrections into account.       These     images    are    displayed   in
Fig.~\ref{fig:xbands}. The maximum   emission is  located  at $\alpha$$^{\rm
EPIC}$(J2000)=05:07:42.40; $\delta$$^{\rm EPIC}$(J2000)=--37:30:46.2   with  an
uncertainty  of 1\farcs5--3\arcsec. The  position of the 2MASS nucleus
is indicated by a  cross. An adaptive   filter was used to smooth  the
images in areas of low signal-to-noise (S/N) ratio, while not smearing
out the emission from point sources.

\begin{figure*}[htb]
\centering
\hbox{\psfig{file=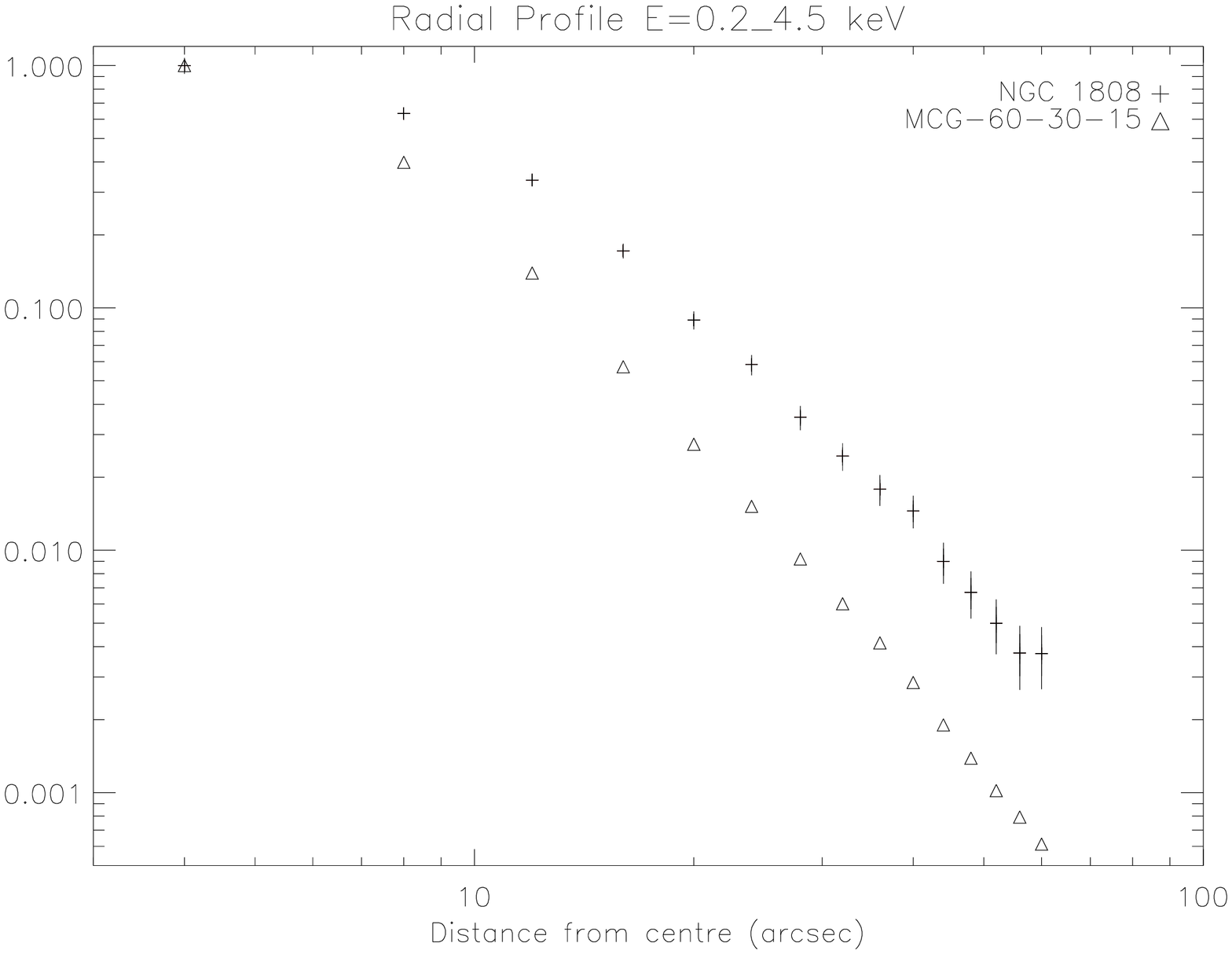,width=80mm,angle=90}\psfig{file=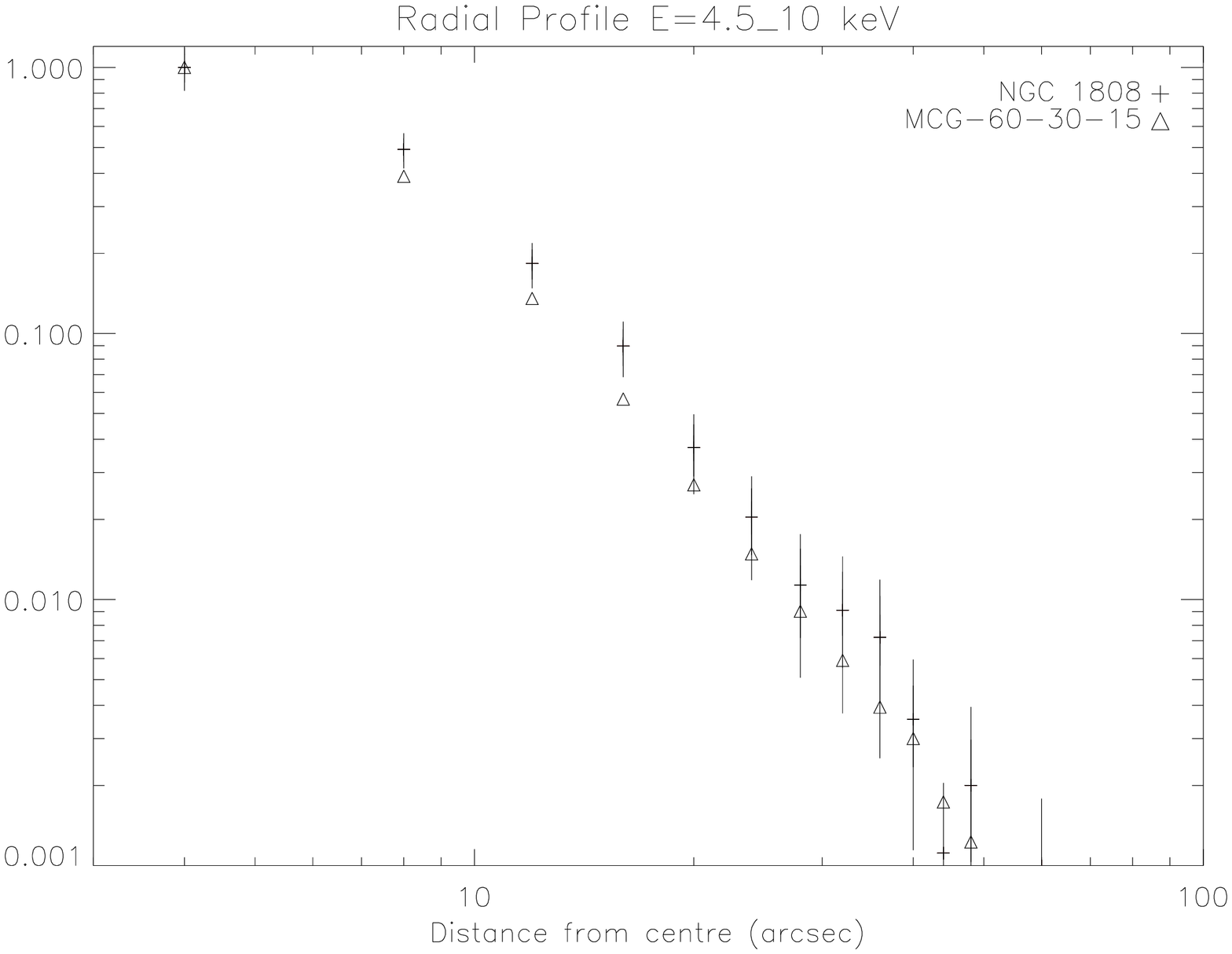,width=80mm,angle=90}}
\caption{Comparison between the radial profiles of NGC\,1808 and the
point-like source MCG--06--30--15 in the 0.2--4.5~keV and 4.5--10~keV
bands.      Y-axis shows    the   normalised  brightness   (normalised
counts/s/px$^2$). The plots  show that  the  emission of  NGC\,1808  is
extended in the 0.2--4.5~keV band and point-like in the 4.5--10~keV
band.}
\label{fig:radial}
\end{figure*}

NGC\,1808 shows extended emission in the soft X-ray images up to 4.5
keV (Fig.~\ref{fig:xbands}). Fig.~\ref{fig:radial}  shows the comparison of  the radial profiles of
NGC\,1808   and  MCG--06--30--15  in two  energy    bands: 0.2--4.5~keV and
4.5--10~keV.  The QSO  MCG--06--30--15 can be  considered  as a  point-like
source in all the \xmm\ energy bands (Kirsch et al. 2005). The comparison shows
that the emission of NGC\,1808 in the lowest energy band is extended
while its radial profile in the 4.5--10~keV band does not
differ from the QSO profile, taking the errors into account.
Therefore, NGC\,1808 can be considered as a point-like source above
4.5~keV. This points out that the origin of the X-ray 
emission is not  due to the compact
central source alone. The elongated X-ray emission of NGC\,1808 follows the
same orientation  as the H$\alpha$ emission (see Sect.~\ref{sect:om}).
The extended morphology is detected at energies  below 4.5 keV, with a
maximum in the 0.5--2.0  keV band (see Fig.~\ref{fig:xbands}). At these
low  energies, thermal emission processes  tracing high-mass SF play an
important role.  Thus, our data suggest that  the stellar  activity in
the nuclear region is  significantly contributing to the observed soft
X-ray emission. With increasing energy,  the central emission  maximum
becomes  more prominent  and  the non-nuclear  sources fade  away.

\begin{figure*}[htb]
\hbox{\psfig{file=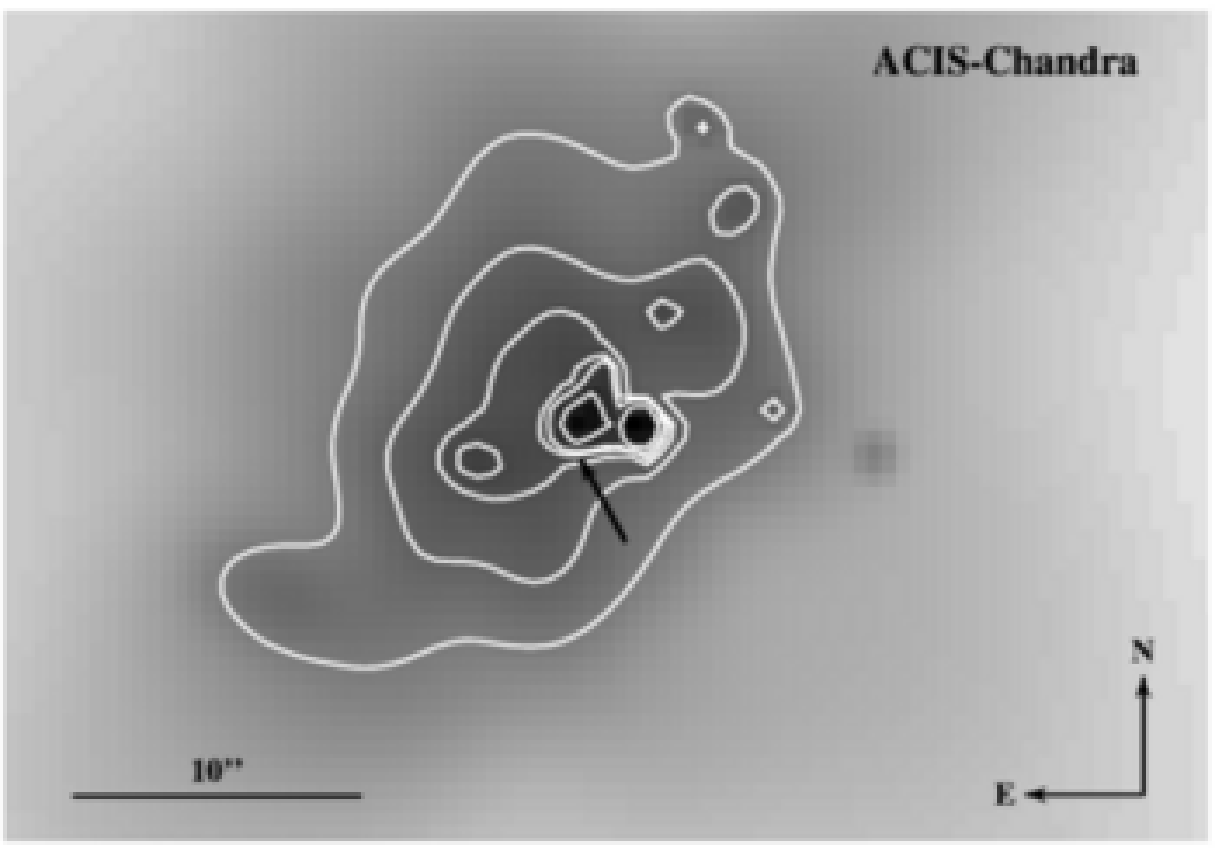,width=8.8cm}\psfig{file=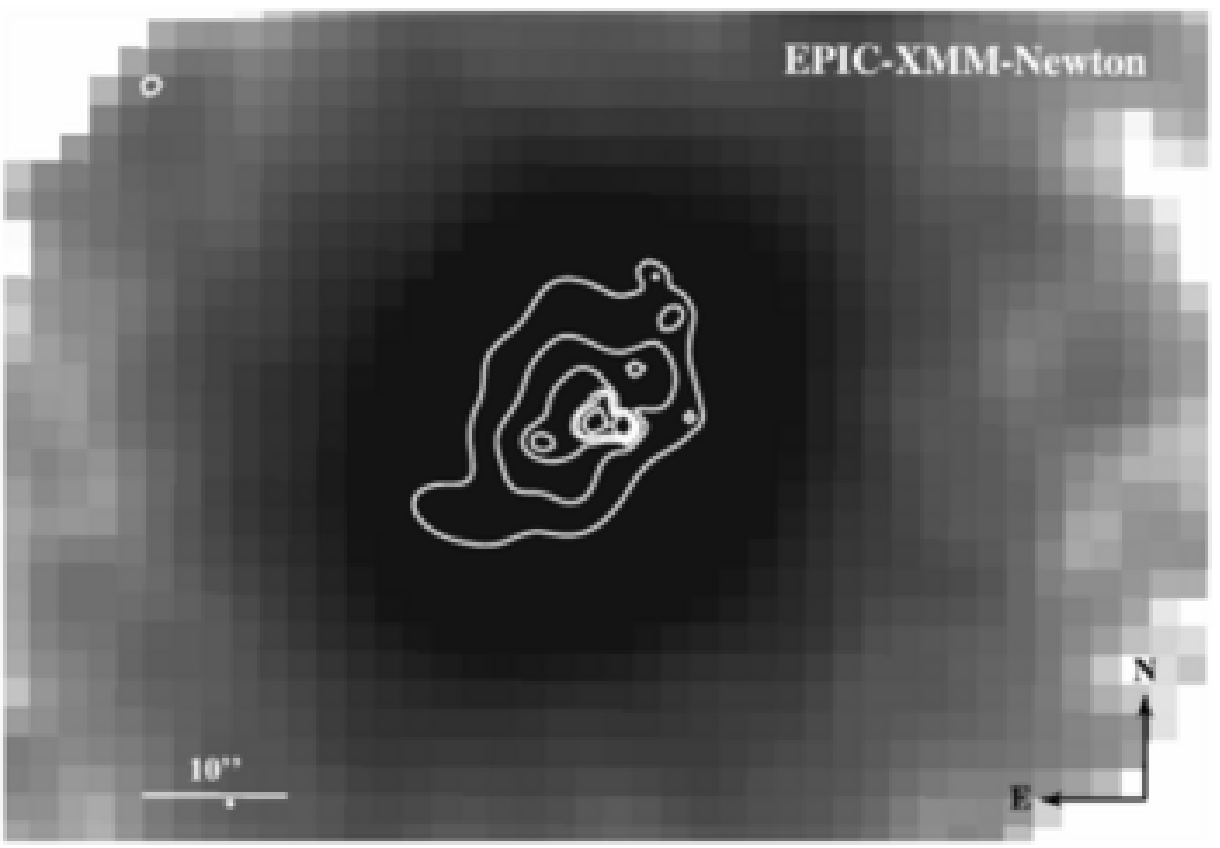,width=8.8cm}}
\caption{NGC\,1808 0.2--12 keV X-ray images taken with the 
\chandra\ ACIS (left panel) and the \xmm\ \epic\ (right
panel)  with the ACIS image  overlaid as contours (contour levels are
2,5,10,15,20, and  50 counts).   The  nuclear  location  from 2MASS  is
indicated   by an  arrow in  the   \chandra\  image.  Both  images are
displayed on  a logarithmic scale with a  threshold of 3$\sigma$  of the
mean background value.  }
\label{fig:xcomp}
\end{figure*}

In order to resolve more details, we have compared our 0.2--12 keV
\epic\ image with the \chandra\ ACIS data. Both were smoothed with an
adaptive filter and are displayed in  Fig.~\ref{fig:xcomp}.  The
\chandra\ ACIS image shows how the nucleus is surrounded by a 
population of discrete sources.   The image (left panel of
Fig.~\ref{fig:xcomp}) shows   the detailed  structure in the central
part of NGC\,1808:  at  least four  point-like  sources are
detected  in the image, apart from  the diffuse emission.  Two of them
are  located at $\sim$4\arcsec\ (212~pc) NW  and SE from the maximum of
the X-ray emission in the \chandra\ image  and have also been detected
in  a \chandra\ HRC  observation.   Zezas  et  al.  (2001)   reported a
luminosity   for   each of  them   of  around  $10^{39}$~erg~s$^{-1}$.

\begin{figure*}[htb]
\centering
\hbox{\psfig{file=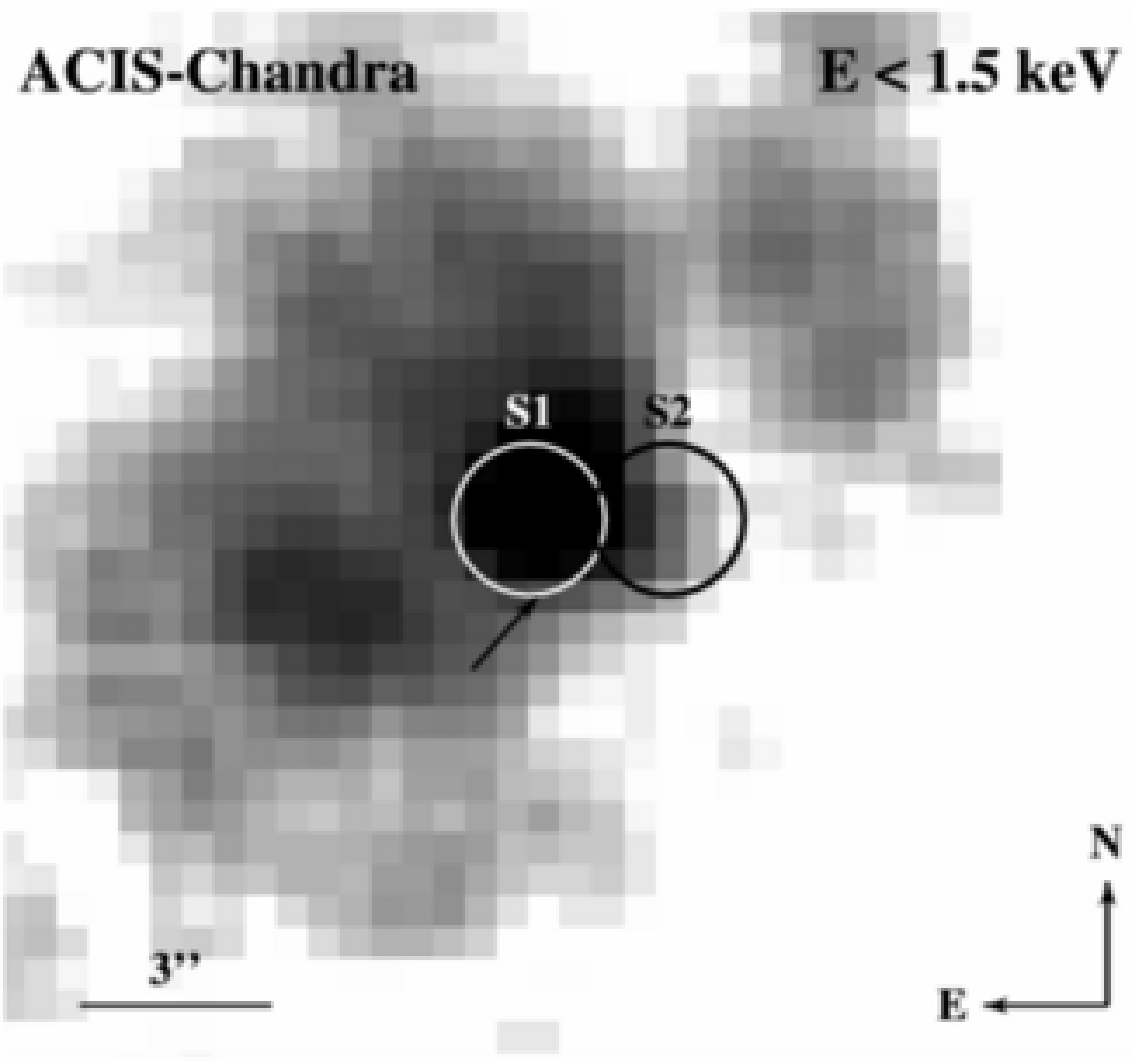,width=8.8cm,angle=0}\psfig{file=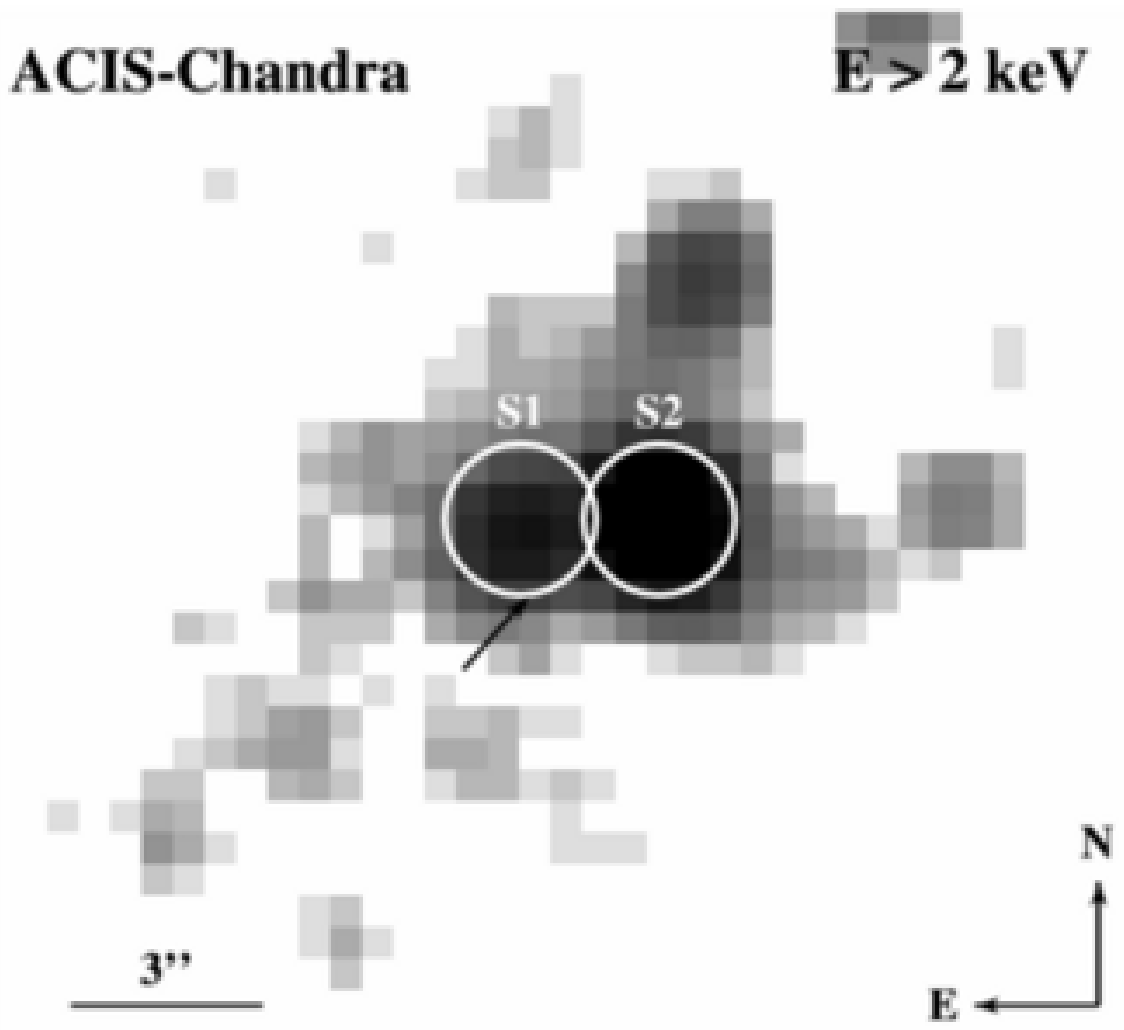,width=8.8cm,angle=0}}
\caption{The \chandra\ ACIS  hard (E$>$2~keV)  and soft (E$<$1.5~keV)
X-ray images  of the two  nuclear  point-like sources.  The  eastern
source, S1, is softer  than the western  one,  S2.  The nuclear
location from 2MASS,  indicated by an  arrow in  the  images, is closer to
source S1. The circles indicate the regions used for spectral extraction
(Sect. \ref{sec:chandraspec}).}
\label{fig:chandra_s1_s2}
\end{figure*}

Although the  nucleus is observed as  a point-like source  in the HRC
image (see Zezas  el al.  2001), the ACIS image reveals the presence
of a double-peak structure (see Fig.~\ref{fig:xcomp}).    The
hard (E$>$2~keV) and soft (E$<$1.5~keV) X-ray images of the two peaks
presented   in  Fig.~\ref{fig:chandra_s1_s2}\ reveal  that  the eastern
source, S1, is significantly  softer than the   western one, S2.  The
locations      of   the      two    peaks   are         $\alpha$$^{\rm
Chandra}_1$(J2000)=05:07:42.35; $\delta$$^{\rm  Chandra}_1$(J2000)=
--37:30:45.8   and $\alpha$$^{\rm  Chandra}_2$(J2000)=05:07:42.19;
$\delta$$^{\rm Chandra}_2$(J2000)= --37:30:45.8 with an uncertainty of
0\farcs6.  The nuclear location from 2MASS is indicated by an arrow in
the \chandra\ image, closer to the eastern source found by \chandra, S1, 
but compatible within the errors with both of them. The remaining two
point-like sources  are  situated on  the ring  of  radio emission  as
imaged by Saikia   et  al.  (1990),  close   to the detected  SNR  and
\hii-regions.

\subsection{X-ray spectral analysis}
\label{sec:spec_ana}

We performed an extensive X-ray spectral analysis of NGC\,1808  
using the \epic\ and \rgs\ instruments on board \xmm, as well as, with  \chandra's 
ACIS camera.

The \xmm\ \epic\ data  were used  to   perform a  spectral
analysis  of NGC\,1808 in  the 0.35--9  keV  energy band with moderate
resolution (80\,eV at  1\,keV). In order  to search for differences in
the  X-ray spectrum   of  the nucleus and  its   surroundings, we
defined three circular regions and one annulus. The circles have radii
of  16\arcsec\ (850\,pc), 35\arcsec\ (1.9\,kpc), and 2\arcmin\ (6.4\,kpc),
while the annulus  has  external and  internal radii  of 20\arcsec\ and
50\arcsec\ (1.0--2.7  kpc),  respectively  (excluding one extra-nuclear
source). Background regions  were selected  to be on the same CCD
as   the     source   but   far     enough    to   prevent    source
contamination.   Figure~\ref{fig:spectra} shows  the   spectra  of  each
region.  We simultaneously  analysed the \pn\ and
\rgs\ spectra, and finally we also studied the \chandra\ spectra of the two
nuclear   point-like      sources          S1  and      S2       (see
Fig.~\ref{fig:chandra_s1_s2}),  both located within the 16\arcsec\ inner
region extracted in the \xmm\ analysis.

In   order    to apply the    modified $\chi^2$  minimization
technique in the spectral analysis,  all  \epic\ spectra were grouped
such that each spectral bin contains at  least 50 counts.  The
\chandra\ spectra were grouped with at least 20 counts in each
bin. The spectra   were  analysed  using  XSPEC v.11.3.0  (Arnaud
1996).

The  spectra  of the  35\arcsec\ and 2\arcmin\ regions are very similar, 
both in shape and intensity (see Fig.~\label{fig:spectra}). Only the spectrum of the 16\arcsec\ region, 
although in good agreement with the two previous ones above 2 keV, 
is clearly weaker in the soft band.  

\begin{figure}[htb]
\vspace{5pt}
\psfig{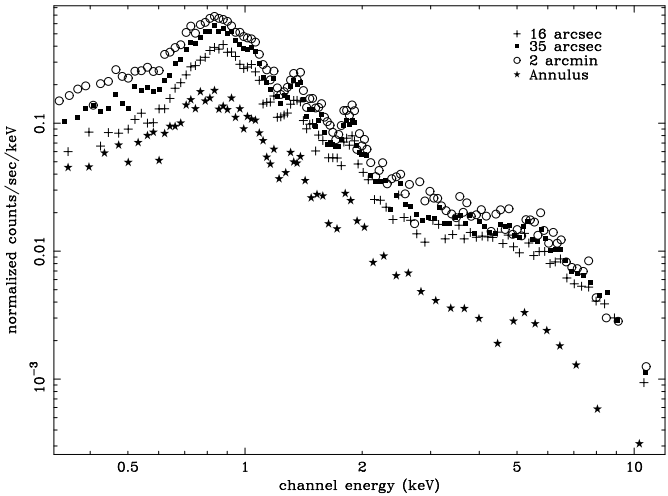}
\caption{EPIC spectra of the three circular regions and the
annulus (see text for further details).
}
\label{fig:spectra}
\end{figure}

\subsubsection{\epic\ spectrum of the inner region}
\label{par:epicspec}

\begin{figure}[htb]
\psfig{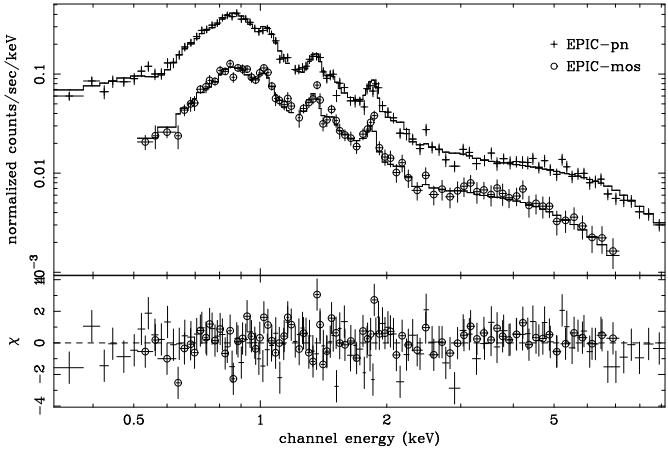}
\caption{EPIC spectrum, best fit model, and residuals of the
r=16\arcsec\ inner region. 
}
\label{fig:nucleus}
\end{figure}

The \epic\ spectrum   of  the inner circular  region  of  NGC\,1808 is
displayed  in  Fig.~\ref{fig:nucleus}.   The  extraction  region has a
radius of 16\arcsec\  (850~pc), centred  on the  maximum peak of   the
X-ray emission.  MOS1  and  MOS2 spectra  and  response matrices  were
combined    by    merging these    to  maximize   the  signal-to-noise
level.  Subsequently,  the \pn\  and  combined MOS  spectra were
fitted simultaneously.
Figure~\ref{fig:nucleus} shows the 0.35--9\,keV spectrum, the best 
fit model with a statistical quality of $\chi_{\nu}^2=0.99$ for 174
degrees of freedom (dof), and the residuals.

Simple    models  as  a  power   law,    a Raymond-Smith (Raymond and
Smith, 1976 ),   mekal (Mewe, Lemen,
\& van den Oord 1986), and
bremsstrahlung  components provide  unacceptable  fits.  The  best fit
model includes,  apart  from the fixed Galactic  foreground absorption
with  a column density  of $N_H=3.23\times10^{20}$\,cm$^{-2}$ (Dickey
\& Lockman, 1990), a power
law  with  index $\Gamma   =   0.79^{+0.07}_{-0.06}$  and a   mekal thermal
component   with a   temperature of  kT=0.580$^{+0.019}_{-0.016}$  keV
absorbed          by      a      Hydrogen          column           of
N$_H$=2.0$^{+0.3}_{-0.4}\times10^{21}\,$cm$^{-2}$.  The value of
$\chi^2$ is unacceptable, but an  intrinsic
absorption  to  the power law component  does  not improve the  fit; the
inferred upper limit     for the    Hydrogen column  density is
$8\times10^{20}\,$cm$^{-2}$. The fit only improved, in terms of
$\chi^2$, when some abundances were allowed to differ from the solar
values.The  abundances of Ne,  Mg, Si, and
Fe,    left    free     in   the    fits,   are   1.6$^{+1.0}_{-0.5}$,
2.2$^{+1.2}_{-0.5}$,  2.1$^{+1.0}_{-0.5}$,  and   0.63$^{+0.2}_{-0.11}$
times the   solar  value ($Z_\odot$),  respectively.  Fits  with other
thermal models such as   Raymond-Smith or bremsstrahlung emission lead  to
higher values of $\chi^2$. No Fe~K$\alpha$ line was significantly
detected. An  upper  limit of   80~eV was  calculated  for  the
equivalent width of  a narrow iron line.  Table~\ref{tab:models} shows
the goodness of the fit together with the values and uncertainties for
each free parameter.

\begin{table*}[htb]
\caption{Parameters and goodness of the best fit models for the r=16\arcsec, r=62\farcs5 circular
regions, and the 16\farcs5--35\arcsec\ annular region.}
\label{tab:models} 
\begin{tabular}{lllllllll}
\hline
{\bf Region} & N$_H$ & $\Gamma$ & A$^\dag$$_{\rm pl}$ & kT$_I$ & A$^\ddag$$_{\rm I}$ & kT$_{II}$ & A$^\ddag$$_{\rm II}$ & Abundances \\
 & $10^{21}\,cm^{-2}$ & & $10^{-5}$& keV & $10^{-4}$ & keV & $10^{-4}$ &   \\ 
\hline

16\arcsec\ (850 pc) & $2.0^{+0.3}_{-0.4}$ & $0.79^{+0.07}_{-0.06}$ & $6.00^{+0.7}_{-0.6}$ & 0.580$^{+0.019}_{-0.016}$ & $3.7^{+0.9}_{-1.2}$ &  --  &  --  &  Ne $1.6^{+1.0}_{-0.5}$ \\
 &    &     &    &    &    &    &    & Mg 2.2$^{+1.2}_{-0.5}$ \\

  &    &     &    &    &    &    &    & Si 2.1$^{+1.0}_{-0.5}$ \\ 
$\chi^2=172$ for 174 dof     &     &    &    &    &    &     & &  Fe 0.63$^{+0.2}_{-0.11}$ \\ \hline
16\farcs5--35\arcsec\ (875--1900~pc) & 2.0 (f)& 0.79(f) & $0.93\pm0.08$ & 0.58 (f) & $1.1\pm0.4$ & $0.096^{+0.015}_{-0.008}$ & $2.5^{+5}_{-1.5}$ & Ne $0.6^{+0.4}_{-0.3}$\\
 &    &     &    &    &    &    &    & Mg $
1.0^{+0.5}_{-0.4}$ \\
  &    &     &    &    &    &    &    & Si $1.2^{+0.5}_{-0.4}$ \\
$\chi^2=135$ for 88 dof      &    &    &    &    &     & & & Fe $0.37^{+0.12}_{-0.04}$ \\ \hline
62\farcs5 (3.3~kpc)& 1.87$\pm$0.03 & 0.79(f) & $6.5\pm0.2$ & 0.572$^{+0.018}_{-0.04}$ & $4.8^{+0.7}_{-1.1}$ & $0.11^{+0.03}_{-0.02}$ & $4^{+15}_{-3}$ & Ne $1.5^{+0.5}_{-0.4}$ \\
&    &     &    &    &    &    &    & Mg 1.9$^{+0.6}_{-0.4}$ \\
  &    &     &    &    &    &    &    & Si 1.9$^{+0.7}_{-0.5}$ \\
 $\chi^2=222$ for 222 dof &    &    &    &    &     &  & &  Fe 0.51$^{+0.11}_{-0.08}$ \\ \hline


\end{tabular}
\\
The best model for 16\arcsec\ region is phabs$_{\rm Gal}$(zpow+zphabs*vmekal)  and the best fit model for 16\farcs5--35\arcsec\ 
and 62\farcs5 regions is phab$_{\rm Gal}$(zpow+zphabs*(vmekal$_I$+vmekal$_{II}$). Normalisation in units of: ($^\dag$)~photons/keV/cm$^{2}$/s at 1 keV
and ($^\ddag$)~in $\frac{10^{-14}}{4 \pi D^2(1+z)^2}\int n_e n_H dV$.
\end{table*}

\subsubsection{The \chandra\ spectra of sources S1 and S2}
\label{sec:chandraspec}

The ACIS image in Fig.~\ref{fig:xcomp} shows a double-peak nuclear
structure surrounded by a few point-like sources inside the
inner, r=16\arcsec, region defined for 
\epic\ and analysed in the previous sub-section. We analysed the standard extracted spectra
obtained with {\tt psextract} of the two circular, r=1\farcs2, regions
S1 and  S2   (see  Fig.~\ref{fig:chandra_s1_s2}). The background   was
extracted  from  a clean circular region   with a radius of 10\arcsec\
located  $\sim$1\arcmin\ North-East   from the  sources  in the   same
CCD. For the  two  regions, models with  a single  component were
rejected due to the high $\chi^2$ value obtained.

\begin{figure*}[htb]
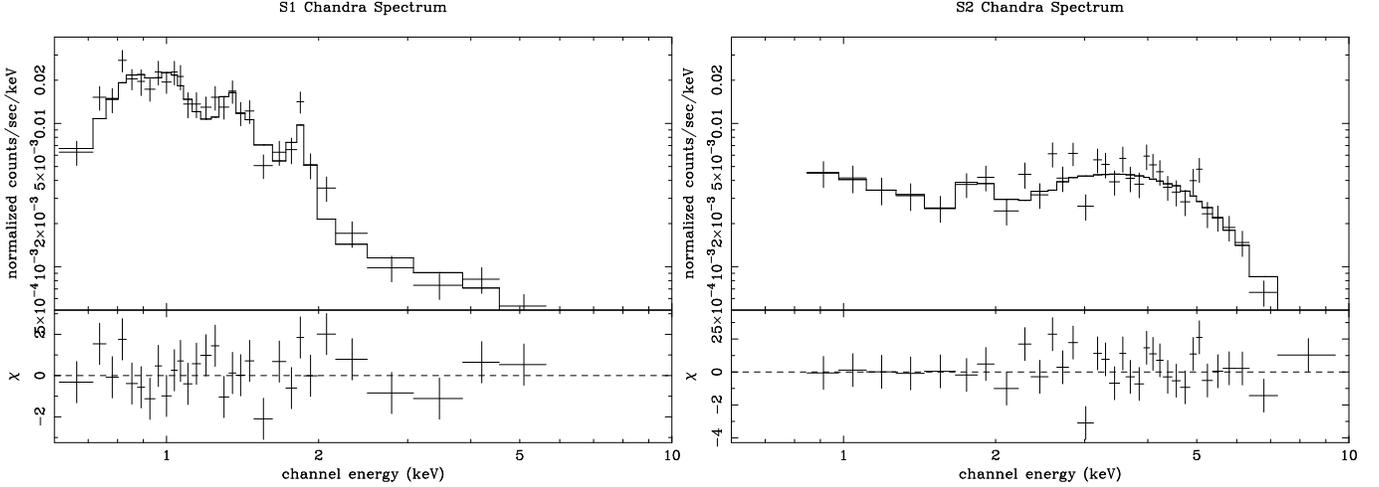


\hbox{\psfig{file=f7a.ps,width=9cm,angle=-90}\psfig{file=f7b.ps,width=9cm,angle=-90}}
\caption{ \chandra\ spectra of source S1~(a) and source S2~(b), the
corresponding best fit
models, and the residuals. Note that the y-axis scale is the same on
both plots to allow comparison of the spectra of both sources.}
\label{fig:spec_chandra}
\end{figure*}

The best  fit  model obtained for the  observed  spectrum of  the hard
source, S2, includes an absorbed mekal component and an absorbed
power  law. The values of  the parameters and  the goodness of the fit
are presented in Table~\ref{tab:models_chandra} and the observed data,
the     best  fit   model,     and   the   residuals   are   shown   in
Fig.~\ref{fig:spec_chandra}b.  It is worth  noting the high values  of
the absorbing column densities for both components, N$_{\rm
{H}}=(2-3)\times10^{22}$~cm$^{-2}$.  The index  of the power law found
is $\Gamma=1.2\pm0.3$, and the temperature of  the thermal component is
kT=$0.087^{+0.005}_{-0.006}$~keV.    The  thermal component  in  S2 is
significantly colder than the one detected in the \epic\ spectrum of
the inner region, with
kT=0.58$\pm0.02$~keV.  The  fit does  not  improve by adding  an extra
thermal component that could account for the thermal emission observed
in the    inner region of \xmm, kT=0.58~keV.
Neither does it improve by varying the abundances from the solar value.

Figure.~\ref{fig:chandra_s1_s2}  shows  that above   2~keV,  source  S1 is
dimmer than source S2, while below 1.5~keV it is the opposite. Assuming
that the low hard emission of S1 could be due  to contamination of the
harder source,  S2, we tested a  model  which includes an absorbed
power law  component with photon index  and N$_H$ fixed to  the values
obtained  for the best  fit model of source  S2. In order to model the
soft range, we  considered  an  absorbed mekal thermal model  with
free    temperature.      The   resulting  fit     is    unacceptable,
$\chi^2_\nu=1.9$.  The addition  of an  extra thermal component   with
kT=0.087~keV  equal  to   the one   detected in   source  S2 does  not
significantly improve the fit.  The best fit model is obtained with an
unabsorbed   power  law   with  $\Gamma=0.99^{+0.18}_{-0.14}$   and  a
moderately absorbed,  N$_H=5.2^{+1.0}_{-0.7}\times 10^{21}$~cm$^{-2}$,
mekal  component  with a  temperature of kT=$0.58^{+0.07}_{-0.09}$~keV
and the abundances of  Ne, Mg, Si, and  Fe fixed to the values obtained
in the fit of the \epic\ spectrum of the inner region. If an absorption 
component above the Galactic value is applied to the power law, the 
value of N$_H$ for that additional component is compatible with zero. 
The observed data, the  best fit model, and  the residuals are shown in 
Fig.~\ref{fig:spec_chandra}a.

\begin{table*}[htb]
\caption{Parameters and goodness of the best fit models for sources S1 and S2 observed with \chandra.}
\label{tab:models_chandra} 
\begin{tabular}{llllllll}
\hline
{\bf Region} & N$_H$(pwlw) & N$_H$(mekal) & $\Gamma$ & A$^\dag$$_{\rm pl}$ & kT & A$^\ddag$ & Abundances \\
& $10^{22}\,cm^{-2}$ & $10^{22}\,cm^{-2}$  & & $10^{-5}$ & keV &  &   \\ 
\hline
{\bf Source S1} & -- & $0.52^{+0.10}_{-0.07}$ & $0.99^{+0.18}_{-0.14}$ & $0.7\pm0.4$ & 0.58$^{+0.07}_{-0.09}$
&  $9^{+3}_{-2}\times10^{-5}$ & Ne 1.6(f)\\
& & & & & & & Mg 2.2(f)\\
& & & & & & & Si 2.2(f)\\
$\chi^2=30$ for 25 dof & & & & & & & Fe 0.63(f)\\ \hline

{\bf Source S2} & $3.1^{+0.8}_{-0.7}$ & $2.0^{+0.3}_{-0.4}$ & $1.2\pm0.3$ & $6^{+7}_{-3}$ & 0.087$^{+0.005}_{-0.006}$ & $2^{+6}_{-2}$ & --\\
$\chi^2=41$ for 28 dof &\\
\hline
\hline
\end{tabular}
\\
The best models for regions S1 and S2 are phabs$_{\rm Gal}$(zpow+zphabs$_{\rm mekal}$*vmekal) and phabs$_{\rm Gal}$(zphabs$_{\rm pwlw}$*zpow+zphabs$_{\rm mekal}$*mekal), respectively. Normalisation in units of: ($^\dag$)~photons/keV/cm$^{2}$/s at 1 keV
and ($^\ddag$)~in $\frac{10^{-14}}{4 \pi D^2(1+z)^2}\int n_e n_H dV$.
\end{table*}

\subsubsection{Simultaneous analysis of \pn\ and \rgs\ spectra}

In order to study the nuclear emission of NGC\,1808 in greater detail, 
we took advantage of the high spectral resolution of the \rgs\ data 
(0.35--2.5 keV).

\begin{figure*}[htb]
\psfig{file=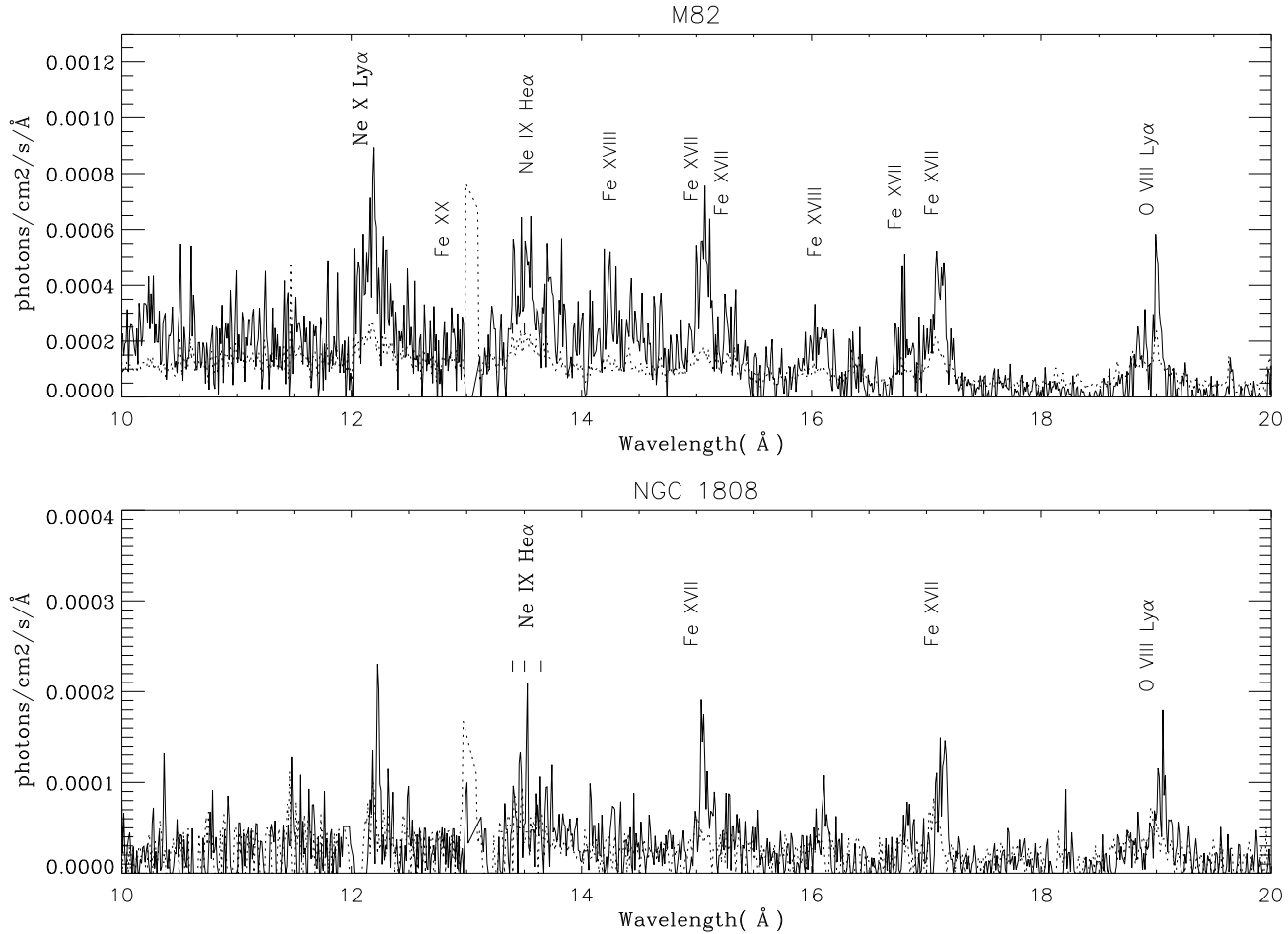,width=176mm,angle=90}
\caption{\rgs\ {\it fluxed} spectra of M\,82 and NGC\,1808, upper and lower
panel, respectively.  The dotted line  corresponds to the errors. The
identification of M\,82 lines is from Reads
\&~Stevens (2002).}
\label{fig:rgs_n1808_m82}
\end{figure*}

Figure~\ref{fig:rgs_n1808_m82} compares the combined {\it RGS1} and {\it
RGS2}  spectrum  of  NGC\,1808 with   that  of M\,82, a  proto-typical
starburst galaxy.  Both spectra were  generated with the SAS task
{\tt rgsfluxer} applying the standard extraction techniques.
The {\it RGS} spectrum of NGC\,1808 shows no continuum emission above 
the noise level, but it does exhibit a number of emission lines. 
Figure~\ref{fig:rgs_n1808_m82} shows that these have wavelengths and relative 
intensity ratios very similar to the strongest lines identified in 
M\,82 (Read and Stevens, 2002). 

We determined the wavelength of the emission lines observed 
with the {\it RGS} by fitting Gaussian profiles to them. The 
redshift-corrected locations of the fitted Gaussians are 
less than 0.02~\AA\ from the theoretical value of the lines 
that we have identified. The fitted line widths are within the range of 
0.02--0.04~\AA,  compatible with the instrumental resolution 
of the {\it RGS}. The detected lines are thus identified as: the Ly$\alpha$ 
emission  line from O\,\ion{VIII}, transitions of He-like Ne\,\ion{IX}\ 
ions, and Fe~L emission lines from Fe\,\ion{XVII}\ and Fe\,\ion{XVIII}. 
The weaker lines  visible in  the M\,82 spectrum were not detected in 
NGC\,1808, most probably due to the lower S/N.
The similarity in the line ratios suggests that there is a common 
origin of the soft X-ray emission in both galaxies, i.e. thermal 
emission from a hot, extended gas component, as shown for M\,82 
(Read and Stevens, 2002). This result confirms the detection of the 
starburst in NGC\,1808 in our data, which dominates the soft X-ray 
emission.

\begin{figure}[htb]
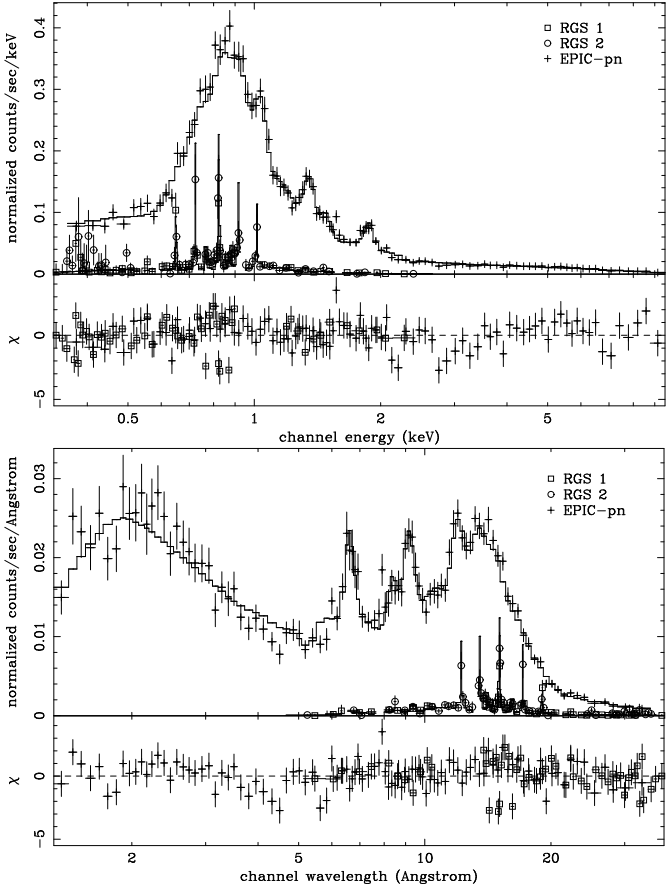

\psfig{file=f9a.ps,width=88mm,angle=-90}
\psfig{file=f9b.ps,width=88mm,angle=-90}
\caption{High-resolution {\it RGS1}+{\it RGS2} and \pn\ spectra in both energy 
(top plot) and wavelength (bottom plot) units, together with the 
best-fitting model which includes a power law and 13 Gaussian 
profiles and the \pn\ and \rgs-1 residuals.}
\label{fig:rgs_pn}
\end{figure}

The \rgs\ spectra of NGC\,1808 and M\,82 in Fig.~\ref{fig:rgs_n1808_m82}
show only a qualitative comparison.  In particular, in the
reduction  process of the  data,  calibration effects affecting  the
line profile were not considered, and  therefore the wings of the lines
are not properly  determined.  The most important  effect of this issue
is that the flux of the lines could be underestimated.  The ratio of fluxes
of the detected lines can be used to derive physical properties of the
gas where the lines originate.  In order  to perform an accurate
study  of the properties  of the emission lines,   we  analysed the
\rgs\ spectrum,  taking the response  of  the detector into account and
therefore avoiding  the    uncertainties explained   before.     Using
XSPEC~v.11.3,   we  simultaneously    analysed  the \rgs\ and  the
16\arcsec\ region \pn\ spectra.  Figure~\ref{fig:rgs_pn}   shows  that  the  soft
X-ray emission  of  NGC\,1808   can   be  explained solely   by    the
contribution of emission lines.  Consequently,  we tried to model
the       {\it  pn-RGS}   spectra    with      an absorbed   (N$_{\rm
H}=1^{+17}_{-1}\times10^{19}$~cm$^{-2}$)            power          law
($\Gamma=1.1^{+0.6}_{-0.5}$                  and               A$_{\rm
{pl}}=9.4^{+0.9}_{-0.7}\times10^{-5}$ photons/keV/cm$^{2}$/s at 1 keV)
which accounts for the  hard  band emission  and   a number of   lines
accounting for the  soft band emission.  The  emission lines were
modeled with Gaussian profiles with widths fixed to their instrumental
value, both in {\it pn} and
\rgs. The  wavelengths were also fixed  to the theoretical values
of the most likely  identifications in order  to reduce the degrees of
freedom.  The best fit model consists of 13 emission lines and a power
law with a value for $\chi^2=307$ for 209 dof.  The energy, intensity,
and        identification    of    each      line    are      shown in
Table~\ref{tab:rgs_pn_fit} and the fit in Fig.~\ref{fig:rgs_pn}.

The line ratios provide a diagnostic of the conditions of the
gas. In  particular, the  ratios between  lines  of iron in  different
degrees of ionization       degrees   allow for    the        best         temperature
diagnostics. Table~\ref{tab:rgs_pn_fit} gives all the ratios relative to
the   strongest  {\it isolated}   line,  i.e.  FeXVII (15.01~\AA).  We
therefore use the observed FeXVII line and compare it with FeXVIII and
others  with  higher    ionization    states to infer      the  plasma
temperature.   We use  the  web version   of   the ATOMDB~v.1.3,  {\it
http://cxc.harvard.edu/atomdb} and    conclude that  a  collisionally
ionized plasma  with kT$\sim$0.4~keV ($\sim5\times10^6$~K) has  a line
spectrum compatible with the observed spectrum of NGC\,1808. This is not
surprising, given our previous results using {\it EPIC} data alone.

\begin{table*}
\centering
\caption{Properties of the identified lines in the simultaneous analysis 
of \rgs\ and \pn\ data.}
\label{tab:rgs_pn_fit} 
\begin{tabular}{l l l l l}
\hline
{\bf Energy } & {\bf Wavelength } & {\bf Intensity} & {\bf I/I$_{{\rm FeXVII}}$} & {\bf Identification} \\
keV & \AA & ph~cm$^{-2}$~s$^{-1}$ \\
\hline\hline
 1.99   &  6.23 & 0.02$^{+0.15}_{-0.02}$ &    0.01&  SiXIV~Ly$\alpha$~6.182~\AA\\
 1.87   &  6.63 & 0.69$^{+0.17}_{-0.15}$ &    0.21&  SiXIII~He$\alpha$~6.65(6.69)6.74~\AA\\
 1.47   &  8.43 & 0.53$^{+0.11}_{-0.19}$ &    0.16&  MgXII~Ly$\alpha$~8.42~\AA\\
 1.34   &  9.25 & 1.09$^{+0.16}_{-0.19}$ &    0.33&  MgXII~He$\alpha$9.17(9.23)9.31~\AA\\
 1.13   & 10.97 & 3.04$^{+0.41}_{-0.38}$ &    0.91&  FeXVII+FeVIII\\
 1.02   & 12.15 & 2.34$^{+0.26}_{-0.24}$ &    0.70&  NeX~Ly$\alpha$~12.134~\AA+FeXVII\\
0.917   & 13.52 & 2.88$^{+0.30}_{-0.29}$ &    0.86&  NeIX~He$\alpha$13.45,13.55,13.70~\AA+FeXVII\\
0.876   & 14.15 & 1.32$^{+0.26}_{-0.26}$ &    0.39&  FeXVIII~14.21,14.37,14.23~\AA\\
0.823   & 15.06 & 3.34$^{+0.30}_{-0.29}$ &    1.00&  FeXVII~15.01,15.26~\AA\\
0.769   & 16.12 & 1.32$^{+0.26}_{-0.26}$ &    0.40&  FeXVIII~16.00,16.07~\AA+OVIII~Ly$\beta$~16.01~\AA+FeXIX\\
0.724   & 17.12 & 2.49$^{+0.33}_{-0.32}$ &    0.75&  FeXVII~16.78,17.05,17.10~\AA\\
0.651   & 19.04 & 1.49$^{+0.34}_{-0.31}$ &    0.45&  OVIII~Ly$\alpha$~18.97~\AA+CaXVII+CaXVI\\
0.565   & 21.94 & 0.17$^{+0.42}_{-0.17}$ &    0.05&  OVII~He$\alpha$~21.6,21.8,22.1~\AA\\
\hline \noalign{ \smallskip} 
\multicolumn{4}{l}{{\it phabs(zphabs*zpow+13zgauss)}  $\chi^2_\nu$=1.47  209  dof }\\
 \noalign{ \smallskip} \hline
\end{tabular}
\end{table*}

\subsubsection{\epic\ spectrum of the non-nuclear extended region}

In   order to   investigate   the  non-nuclear  extended emission   of
NGC\,1808, the  \pn\ spectrum of an  annular region  was extracted
and analysed.  The inner  and the outer  radii, 16\farcs5 (875~pc) and
35\arcsec\ (1.9~kpc), were  chosen to minimize the contribution of
the so-called {\it inner region},  i.e.  r=16\arcsec, and to  exclude
the CCD edges.

\begin{figure}[htb]
\psfig{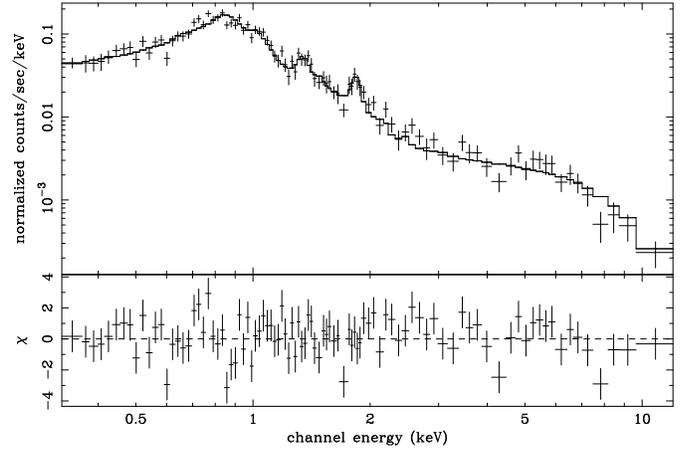}
\caption{\pn\ spectrum, best fit model, and residuals of the
annular region. }
\label{fig:annular}
\end{figure}

For the spectral fit, we have assumed that the emission in this region
is contaminated by the emission from the inner region.  Therefore, the
fitting  model includes a  power law with an  index fixed to the value
obtained in the fit of  the inner region spectrum, i.e. $\Gamma=0.79$,
and  an   absorbed  {\it   mekal}  component   with   N$_H$  fixed  to
2$\times10^{21}\,$cm$^{-2}$ and kT$_{\rm  I}$ fixed to  0.58~keV.  The
normalisations of  both components  were  left free.  An  excess  with
respect to the model is observed at low energies.  This feature can be
well fitted  with  an extra  mekal component  with a temperature
kT$_{\rm II}\simeq0.1$~keV.  The
abundances of Ne,  Mg, Si, and  Fe have been left  free in the fits and
their values  are  compatible within the   errors with the  abundances
obtained for the inner region.  The   values   of  all
parameters  and   the goodness    of    the  fit  can  be    found  in
Table~\ref{tab:models}.   Figure~\ref{fig:annular}  shows the  data, the
best fit model,   and the  residuals.

Comparison of  the inner and  annular regions shows that the annular
spectrum includes  a component not detected  in the  inner region, the
low temperature  gas, which  can  therefore be  associated with  a
non-nuclear, extended emission component.

The results of the analysis confirm our expectation that the spectrum of
the annular region   is contaminated by  some emission from the  inner
region.  This  is especially  true   for the  power  law,  which is 
fainter by a factor  of $\sim$6.5 compared to that in the inner region,  
while the hotter
thermal gas   contribution may be  partially   due to either  extended
emission or discrete sources  in the annulus,  as its normalisation is
only a  factor of 3.4 weaker than in  the inner  region.  Indeed,  the
encircled energy function for EPIC-pn predicts  that about 20\% of the  
total energy emitted by the unresolved source will be imaged between
r=16\farcs5\ and r=35\arcsec\ from the centre of the point spread 
function.

\subsubsection{Integrated \epic\ spectrum of the total galaxy}

A circular region with $r = 62\farcs5$ (3.3~kpc) was analysed. The
extracted \epic\ spectrum contains the emission of the whole galaxy and 
the region coincides with the extraction radius used in the \rosat\ 
analysis (Dahlem et al. 1994; Junkes et al. 1995). 

\begin{figure}[htb]
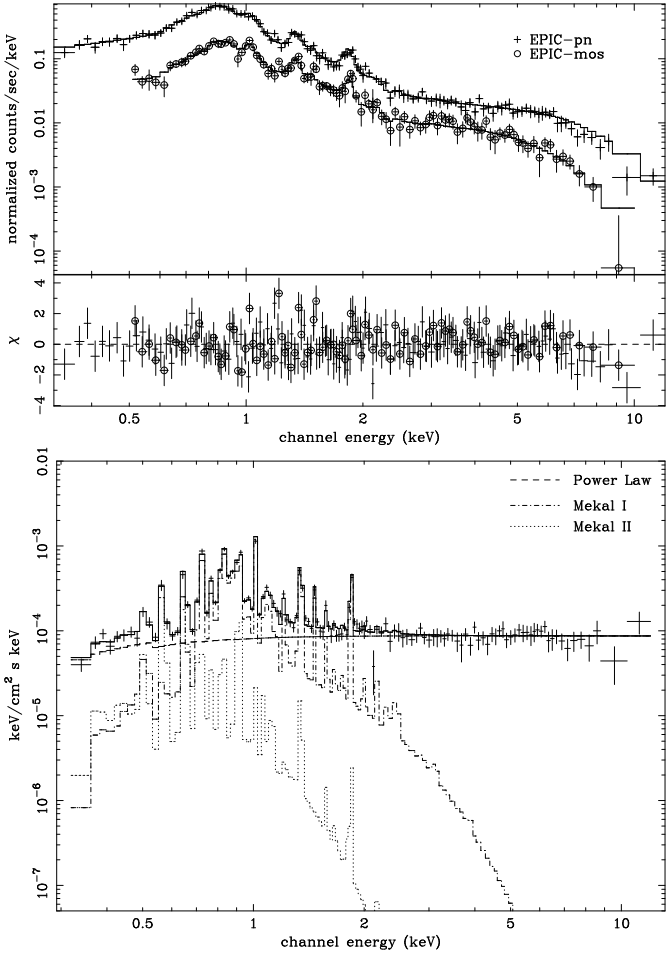

\psfig{file=f11a.ps,width=88mm,angle=-90}
\psfig{file=f11b.ps,width=88mm,angle=-90}
\caption{\epic\ spectrum, best fit model, and residuals of the
r=62\farcs5 (3.3~kpc)  region (top plot) and the best fit model, the
various additive components, and the \pn\ spectrum in $Ef(E)$ (bottom plot). }
\label{fig:galaxy}
\end{figure}

Analogously to  the annular region,  the  components  of the best  fit
model (   $\chi^2_\nu=1.00$ for 222   dof) are  a  power  law and  two
absorbed thermal components.  The values of the various parameters and
the   goodness  of the  fit  can  be  found in Table~\ref{tab:models}.
Figure~\ref{fig:galaxy}  shows  the data,  the  best  fit model,  and the
residuals. In the  lower panel of  Fig.~\ref{fig:galaxy}, we also show
the  best fit  model and  its various additive  components in $Ef(E)$.
The  values of the  parameters suggest that the  emission has the same
origin as for  the annular  region; i.e.  the  power  law and higher
temperature component are associated with the nuclear emission, and the
soft  thermal emission  is  associated  with the non-nuclear  extended
emission.

\subsection{X-ray fluxes and luminosities}

Various absorbed X-ray fluxes  and unabsorbed luminosities, measured
in the three regions   of the \xmm\ analysis  and the two regions of
the \chandra\ analysis, are collated in Table~\ref{tab:lumin}. 

For all  cases, the thermal  components  dominate  in the soft  X-ray
band, while the  power law dominates  in  the 2--10 keV  band. The
innermost part of the galaxy (r=16\arcsec=850~pc) is responsible for
emission in the hard band, while in the soft energy band it contributes 
only  half of the total luminosity of the galaxy ($r=62\farcs5$=3.3~kpc).

\begin{table*}[htb]
\centering
\caption{NGC\,1808 X-ray absorbed fluxes and unabsorbed luminosities}
\label{tab:lumin} 
\begin{tabular}{lllll}
\hline
{\bf Region} &  F$_{0.1-2.4\,keV}$ & F$_{2-10\,keV}$ &  L$_{0.1-2.4\,keV}$ & L$_{2-10\,keV}$  \\
radius & $10^{-13}$\,erg\,s$^{-1}$cm$^{-2}$ & $10^{-13}\,$erg\,s$^{-1}$cm$^{-2}$ & $10^{39}$\,erg\,s$^{-1}$ & $10^{39}\,$erg\,s$^{-1}$ \\
\hline\hline
16\arcsec\ (850~pc) & $5.8^{+1.1}_{-1.2}$ & $11.2\pm0.4$ & $18\pm4$  & $16.1\pm0.6$ \\
\,\,{\it Power law} & $1.95\pm0.17$ & $11.0\pm0.4$ &  $3.2\pm0.3$($\sim 20\%$) & $15.6\pm0.06$ ($\sim 98\%$)\\
\,\,{\it Thermal }  & $3.8^{+1.1}_{-1.2}$ & $2.1^{+0.6}_{-0.7}$ & $15\pm4$($\sim 80\%$) & $0.33^{+0.10}_{-0.11}$ ($\sim 2\%$)\\
\hline

16\arcsec--35\arcsec\ (875--1900~pc)($\dag$)& 1.54$^{+0.16}_{-0.2}$ & 1.78$^{+0.17}_{-0.19}$ & 8.2$^{+1.0}_{-1.4}$ & 2.5$\pm0.5$ \\
\,\,{\it Power law} & 0.33$^{+0.3}_{-0.12}$ & 1.71$^{+0.17}_{-0.16}$ & 0.5$\pm0.2$ ($\sim 5\%$) & 2.4$\pm0.6$($\sim 100 \%$)\\
\,\,{\it Thermal I}  & 1.1$\pm0.02$ & 0.075$^{+0.016}_{-0.002}$ & 4.3$^{+1.1}_{-1.3}$($\sim40\%$) & 0.18$^{+0.2}_{-0.02}$ ($ \sim5 \%$)\\ 
\,\,{\it Thermal II}  & 0.09$^{+0.3}_{-0.05}$ & $-$ & 3.4$^{+1.6}_{-1.2}$ ($\sim55\%$) &
$-$\\ \hline

62\farcs5 (3.3~kpc) ($\dag$) & 6.8$^{+0.6}_{-0.7}$ & 12.1$\pm0.6$ & 26$^{+5}_{-9}$ & 17.2$^{+0.8}_{-0.6}$ \\ 
\,\,{\it Power law} & 2.1$\pm1.4$  & 11.9$\pm0.6$ & 3.3$^{+9}_{-0.3}$ ($\sim 10\%$)  & 16.9$^{+0.7}_{-0.6}$ ($\sim 98\%$)  \\
\,\,{\it Thermal I}  & 4.6$^{+0.6}_{-0.9}$ & 0.24$^{+0.07}_{-0.02}$ & 15$^{+7}_{-3}$ ($\sim 60\%$) & 0.38$^{+0.9}_{-0.08}$ ($\sim2\%$) \\
\,\,{\it Thermal II} & 0.18$^{+1.2}_{-0.11}$ & $-$ &7$^{+7}_{-6}$ ($\sim 30\%$) &
$-$\\ \hline
S1 (65 pc) & 0.63$^{+0.06}_{-0.10}$ & 1.0$^{+0.3}_{-0.5}$ & 3.9$^{+0.3}_{-1.0}$ & 1.5$^{+0.2}_{-0.9}$ \\
\,\,{\it Power law}& 0.22$^{+0.13}_{-0.06}$ & 1.0$^{+0.2}_{-0.6}$ & 0.39$^{+0.8}_{-0.03}$ ($\sim 10\%$) & 1.4$^{+0.7}_{-0.3}$ ($\sim 95\%$)\\ 
\,\,{\it Thermal } & 0.41$^{+0.10}_{-0.06}$ & 0.045$^{+0.2}_{-0.002}$ & 3.5$^{+0.4}_{-0.8}$ ($\sim 90\%$)& 0.074$^{+0.03}_{-0.007}$ ($\sim 5\%$)\\

\hline

S2(65 pc) & 0.27$^{+0.09}_{-0.18}$ & 5.4$^{+0.8}_{-5}$ & $<6\times 10^4$ & 9.2$^{+0.8}_{-7}$\\
\,\,{\it Power law} & 0.20$^{+0.12}_{-0.19}$ & 5.4$^{+0.5}_{-5}$ & $<400$ ($\sim 0\%$) & 9.4$^{+0.9}_{-0.7}$ ($\sim 100\%$)\\
\,\,{\it Thermal } & 0.07$^{+0.07}_{-0.03}$ & 3$\times 10^{-5}$ & $<6\times 10^4$
($\sim 100\%$) & $-$   \\
\hline

16\arcsec\ (850~pc) \chandra & 5.5$^{+0.4}_{-0.9}$ & 9.4$^{+0.6}_{-0.9}$ &
-- & -- \\

\hline
16\arcsec--35\arcsec\ (875--1900~pc) \chandra & 0.5$^{+0.4}_{-0.3}$ &
$<0.23$ & -- & -- \\

\hline
62\farcs5 (3.3~kpc) \chandra & 6.1$^{+0.4}_{-1.2}$ &
9.1$^{+0.4}_{-1.2}$ & 26$^{+2}_{-9}$ & 22.2$^{+2}_{-1.6}$ \\
\hline
62\farcs5 (3.3~kpc) \rosat & & & $\sim140$ & -- \\
\hline
\end{tabular}\\
($\dag$) The label {\it Thermal I} corresponds to the highest temperature
mekal component and the {\it Thermal II} to the lowest.
\end{table*}

In  order to study  the long  term variability  in  NGC\,1808, we 
compared the fluxes  observed with \xmm\ in  April 2002 with the  ones
observed with \chandra\ eight  months later.  To allow comparison, the
spectra of  the nuclear, the annular, and  the inner regions were
extracted from the  \chandra\   ACIS observation.  These  regions  are
equivalent to the ones analysed with \xmm. The spectra of
the inner regions were fitted using  the best fit model obtained
from the  \xmm\ analysis,  i.e.  a  power law  plus an absorbed  mekal
component. The index of the power law, hydrogen column density, and the
temperature  and abundances  of the  mekal  component of the  best fit
model  ($\chi^2=170$ for 137   dof)  are  compatible with the   values
obtained  with  {\it XMM-Newton}.  Therefore,  no spectral variability
was   detected  between both observations.  The   spectrum of the
annular region was also fitted using the  best fit model obtained
with \xmm. The values of the relevant parameters of the best fit model
($\chi^2=45$ for 36  dof) are  again compatible  with the  ones derived
from the \xmm\  observation.

In the inner region, the comparison of the  \xmm\ and \chandra\ values
reveals  no sign of  variability in the  soft  energy range \chandra\
flux  is   only 5\% lower    than  the \xmm\  flux,  well   within the
statistical errors  and calibration uncertainties of both
instruments. In the hard band,  a decrease of 16$\%$ in  flux is observed at a
2.5$\sigma$ level (see Table~\ref{tab:lumin}).

The \chandra\ analysis of the two discrete sources detected within the
r=16\arcsec\ region reveals that source S2 is responsible for $\sim$50\%
of   the  hard  emission from  the    inner \chandra\ region, while the
combined emission  of sources S1 and S2  only  explains $\sim$15\% of
the  observed soft emission in this  inner region. It  is worth noting
that the upper limit of the intrinsic, i.e. absorption-corrected  soft
X-ray  luminosity of  S2  obtained with \chandra, is  more than  three
orders  of  magnitude   higher  than  the   luminosity  of the   whole
galaxy. However, considering the uncertainties of this measurement, it 
is compatible with lower values.   Moreover,  taking  the
absorbed  fluxes into account, no   discrepancy    is found, indicating   that  the
intrinsic   absorption   components  are   responsible   for  the high
luminosity measured.  Although  the  statistical test shows   that the
model is acceptable, alternative scenarios cannot be ruled out.

In the annular region, the \chandra\ flux measurement in the soft
band is significantly lower than those of \xmm. However, 
\chandra\ is less efficient than \xmm\ to detect weak extended emission 
and probably the discrepancy is due to the extended emission not
determined by \chandra.

We  also measured the fluxes and luminosities with \chandra\ within 
an r=62\farcs5 aperture to compare the \xmm\ data with \chandra. The 
\chandra\ spectrum does not require the complexity needed to describe
the \xmm\ spectrum: only a combination of a power law and one mekal
component is sufficient. The
lowest temperature mekal, kT$\sim0.1$~keV,   found with \xmm\  is  not
detected in the \chandra\ spectrum due  to the small effective area of
the instrument at such low  energies. The values  of the parameters of
the two  components included in  the best-fit  model ($\chi^2$=220 for
200 dof) are compatible with the \xmm\ results.

As expected,  taking the results for  the inner region into  account, 
a mild variability in the hard band was  observed between the \xmm\ and
\chandra\ observations, while no significant  variation of the flux was
measured in the soft energy range. This variation is dominated by
the power law component of the inner region.

In     addition   to   the   newly     derived    luminosity   values,
Table~\ref{tab:lumin} also lists the  luminosity measured with \rosat\
(Junkes et al.  1995) in this region. The values indicate an important
decrease in the luminosity  between the  \rosat, \xmm, and \chandra\
observations.                  Comparing            our        results
($L_{0.5-10\,keV}\sim3\times10^{40}$~ergs$^{-1}$)  for the r=62\farcs5
region with the \asca\ observations of February 1994 obtained by Awaki
et al.   1996   ($L_{0.5-10\,keV}=2\times10^{40}$~ergs$^{-1}$), a small
increase in the  luminosity is observed.  The comparison with the
\ginga\ results (October 1990) indicates  a decrease of  the luminosity
from  $L_{2-10\,keV}=5\times10^{40}$~ergs$^{-1}$     for     \ginga\ to
$L_{2-10\,keV}=1.7\times10^{40}$~ergs$^{-1}$  for \xmm.  Nevertheless,
existing  uncertainties in the  relative calibration of various X-ray
detectors  could  explain   the  discrepancy    of  \xmm\ and
\chandra\ with the \asca\ and \ginga\ values, but the luminosity measured
by \rosat\ is 5 times that observed with \xmm\ and \chandra, too high to
be explained by calibration uncertainties.

Short-term variability during the \xmm\ observation was not detected.

\subsection{{\it OM} optical/UV data and H$\alpha$ imagery}\label{sect:om}

\begin{figure*}[htb]
\centering
\hbox{\psfig{file=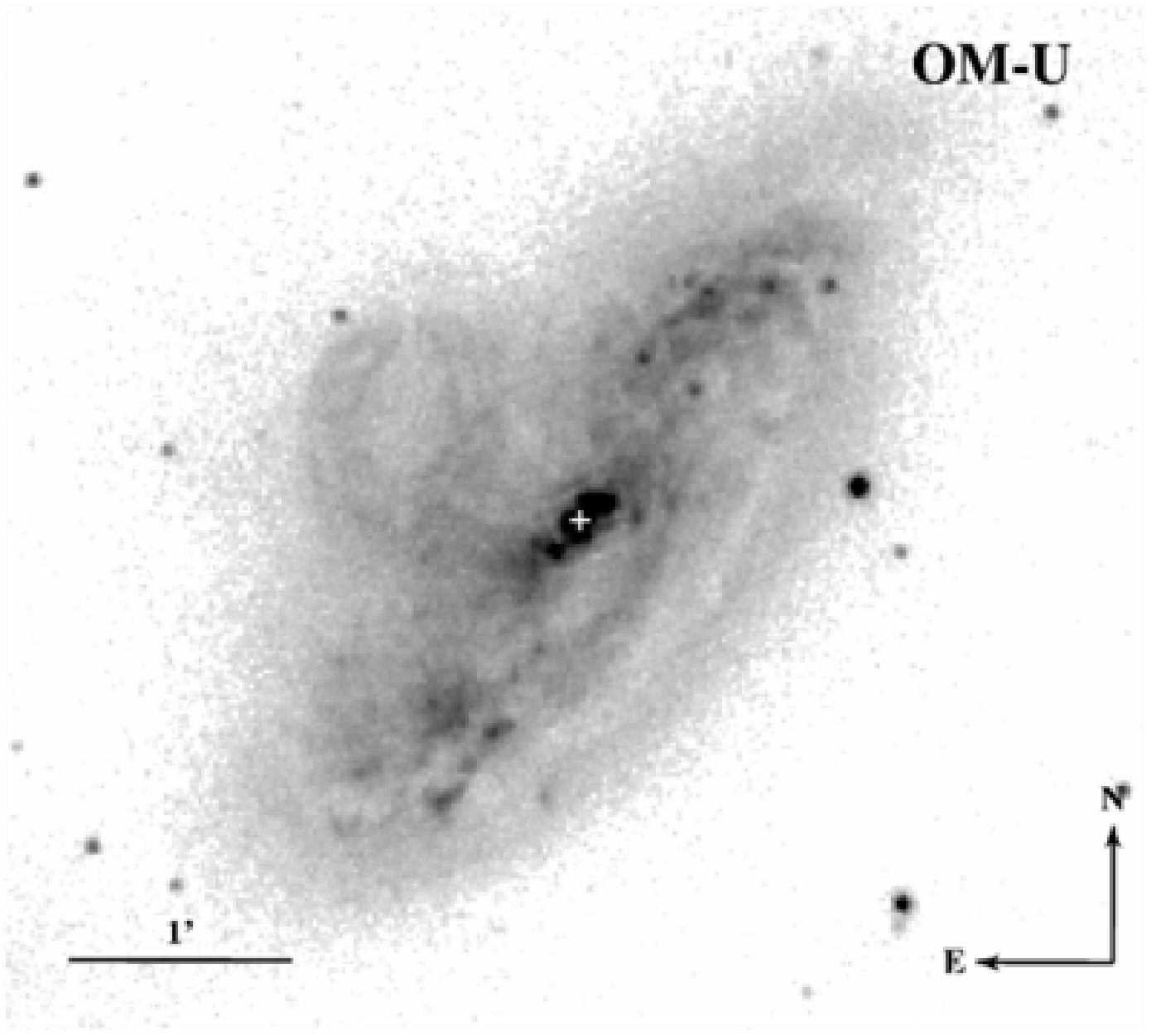,width=9cm,angle=0}\psfig{file=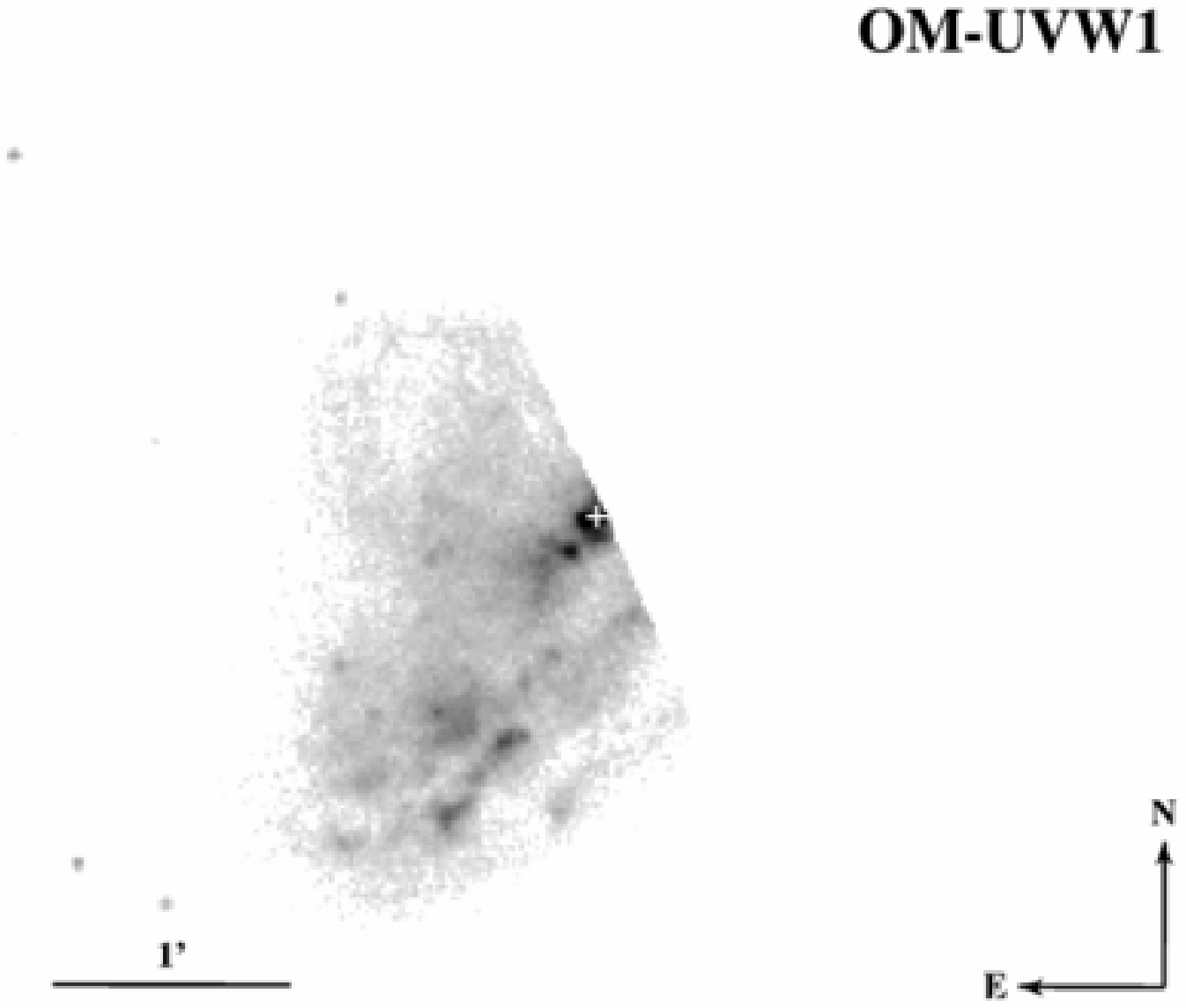,width=9cm,angle=0}}
\hbox{\psfig{file=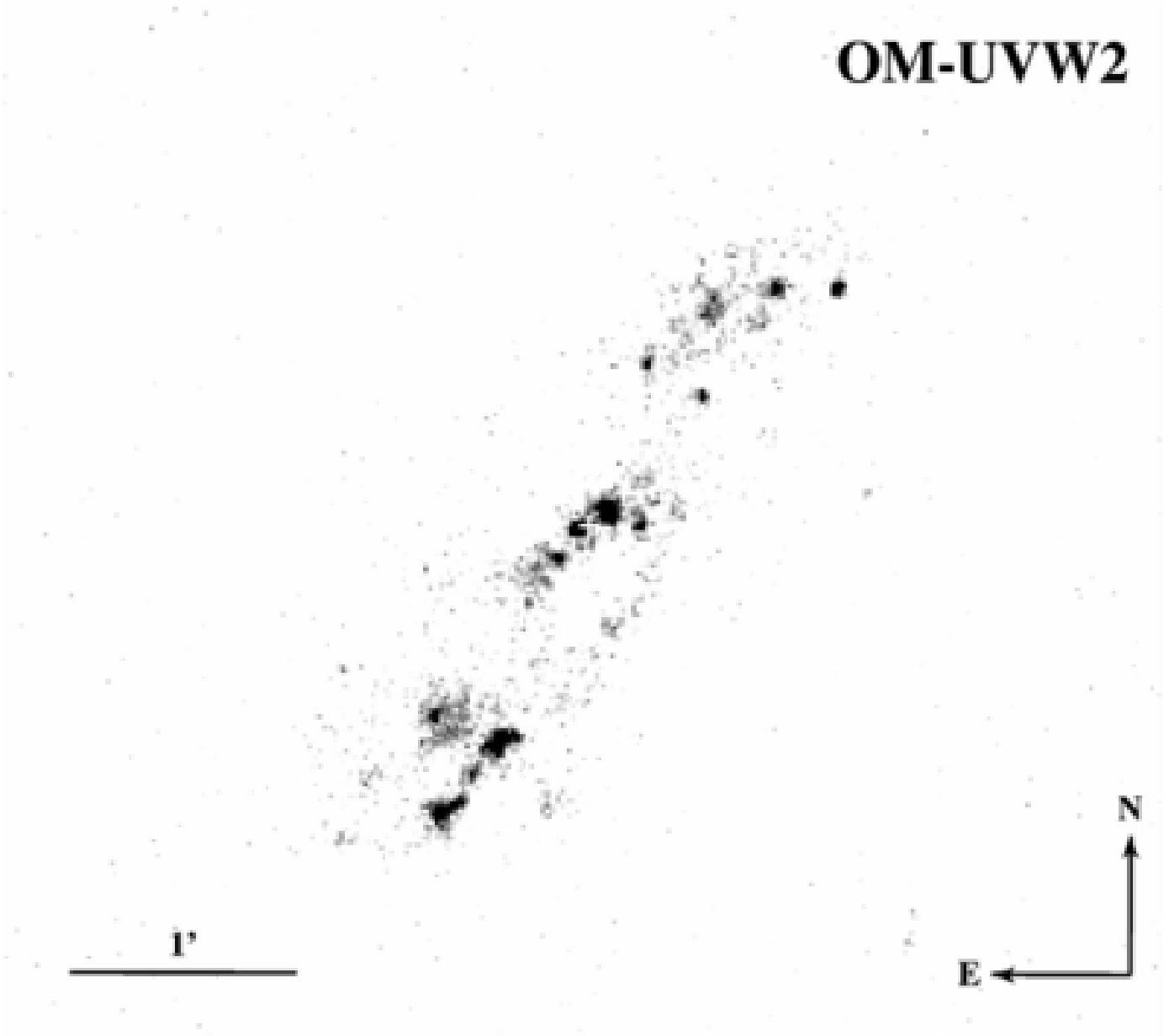,width=9cm,angle=0}\psfig{file=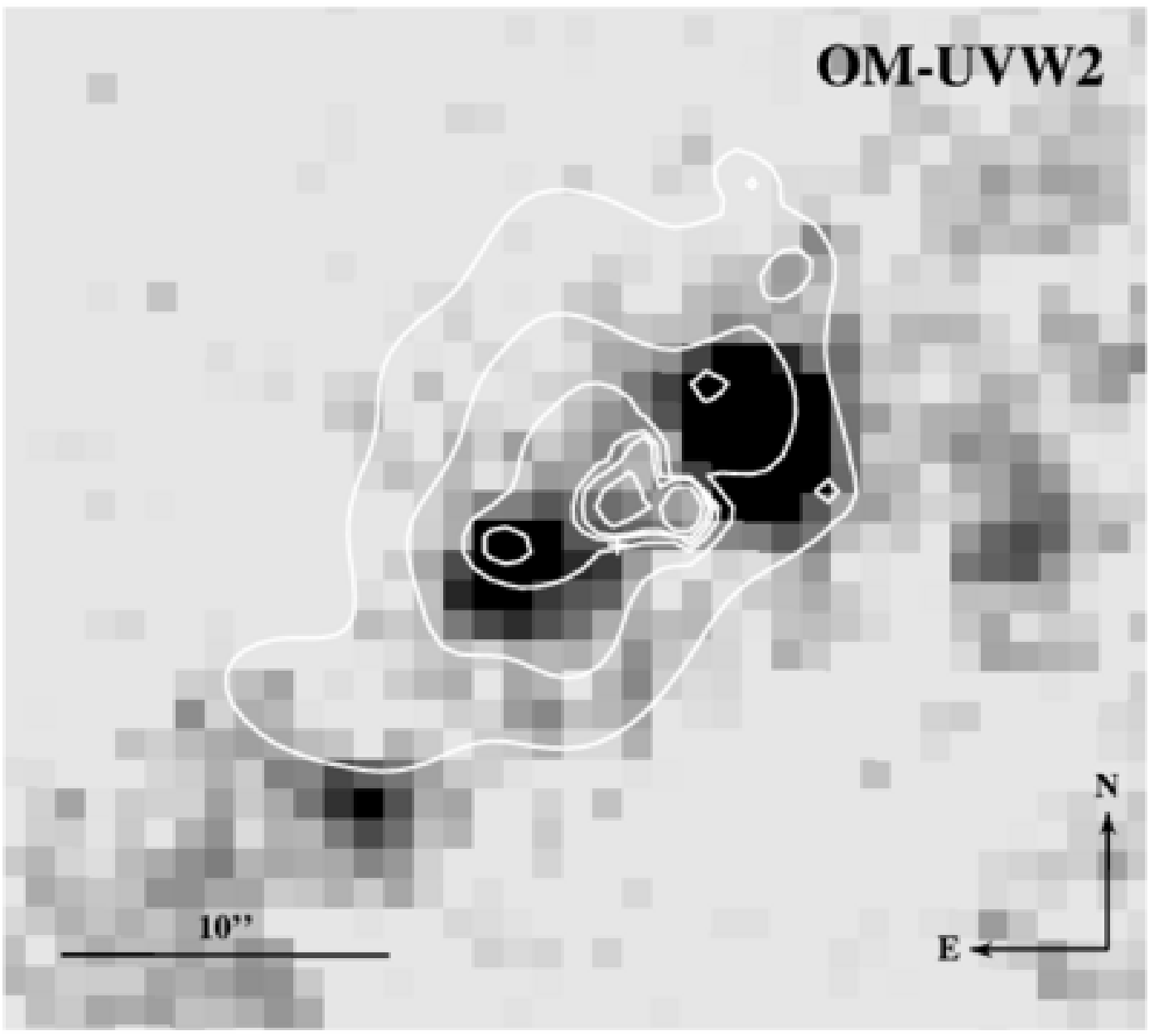,width=9cm,angle=0}}
\caption{{\it OM} UV images of NGC\,1808. From left to right and from
top to bottom: U filter (350~nm), UVW1 filter (291  nm), UVW2 (212 nm)
images, and the UVW2 central region of NGC\,1808 with the overlay of the
\chandra\ contours. The location of the nucleus from 2MASS is marked with a
cross.  The elliptical ring  eastwards of   the galaxy  in U  and UVW1
images is an artefact due to the light scattered within the
\om\ detector. All images are displayed in logarithmic
scale with a threshold of 3$\sigma$ of the mean background value.}\label{fig:uv}
\end{figure*}

{\it OM} UV images of NGC\,1808 with the U (350 nm), UVW1 (291 nm) and
UVW2 (212    nm) filters  were   taken  in  parallel   with  the  X-ray
observations. Due  to technical problems,  part of  the UVW1 image was
lost. All  images are displayed in Fig.~\ref{fig:uv}.   The U and UVW2
images show a double-peak structure in the  centre of the galaxy. The
locations   of    the  peaks     in  UVW2    image     are   $\alpha^{\rm
OM}_1$(J2000)=05:07:41.97;  $\delta$$^{\rm  OM}_1$(J2000)=--37:30:43.3    and
$\alpha$$^{\rm OM}_2$(J2000)=05:07:42.62; $\delta$$^{\rm OM}_2$(J2000)=--37:30:48.1
with uncertainties of $\sim$2\arcsec. Figure~\ref{fig:uv}.c shows the
central region of the UVW2 image with the \chandra\ contours
overlaid. The locations of the two point-like sources detected by
\chandra\ (at $\sim$4\arcsec\ from the nucleus and aligned in the NW and
SE direction) coincide with the positions of the UV peaks observed in
the \om\ image. However, there is no UV emission associated with any of
the two bright X-ray sources at the very nucleus of the galaxy.

\begin{figure*}[htb]
\hbox {\psfig{file=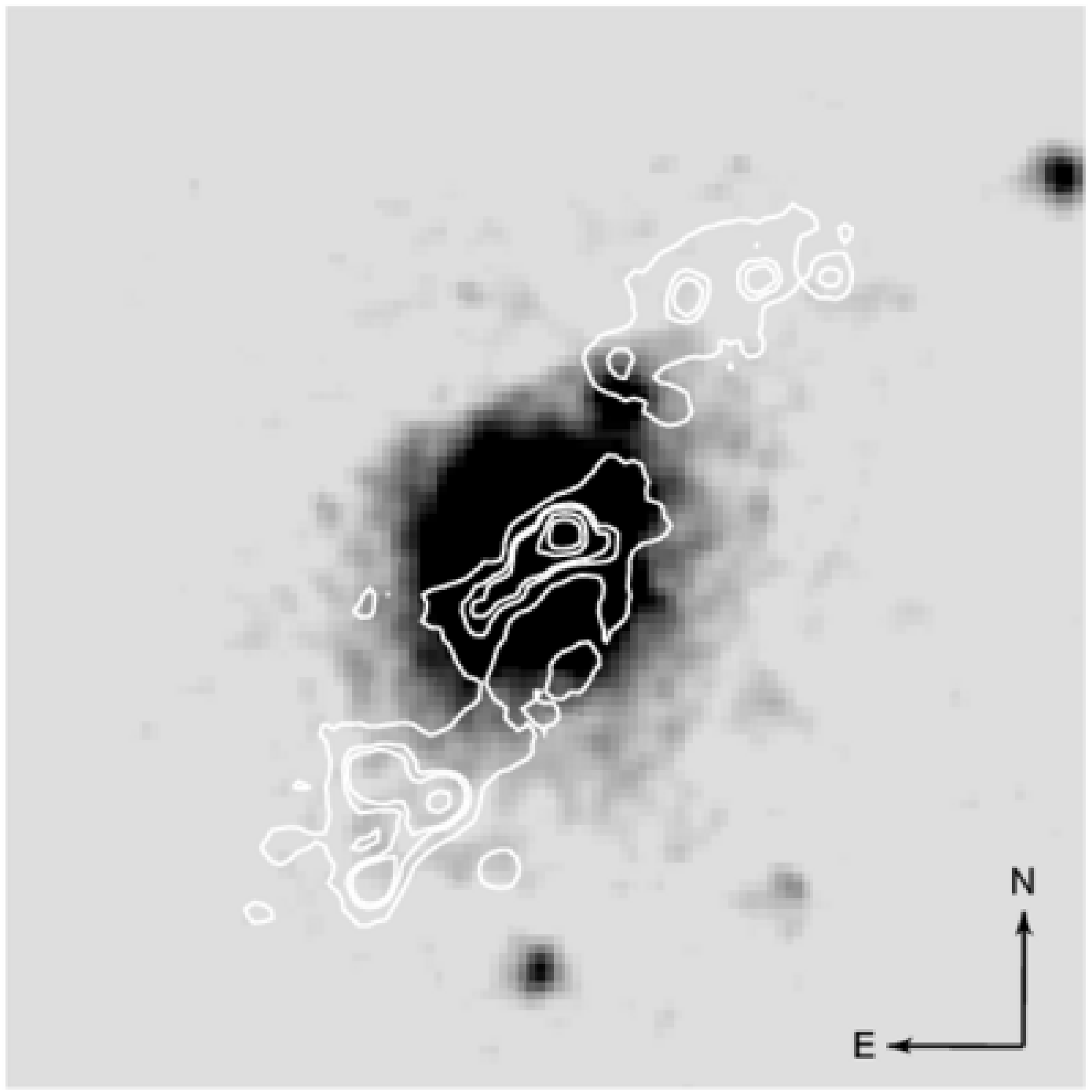,width=9cm,angle=0} \psfig{file=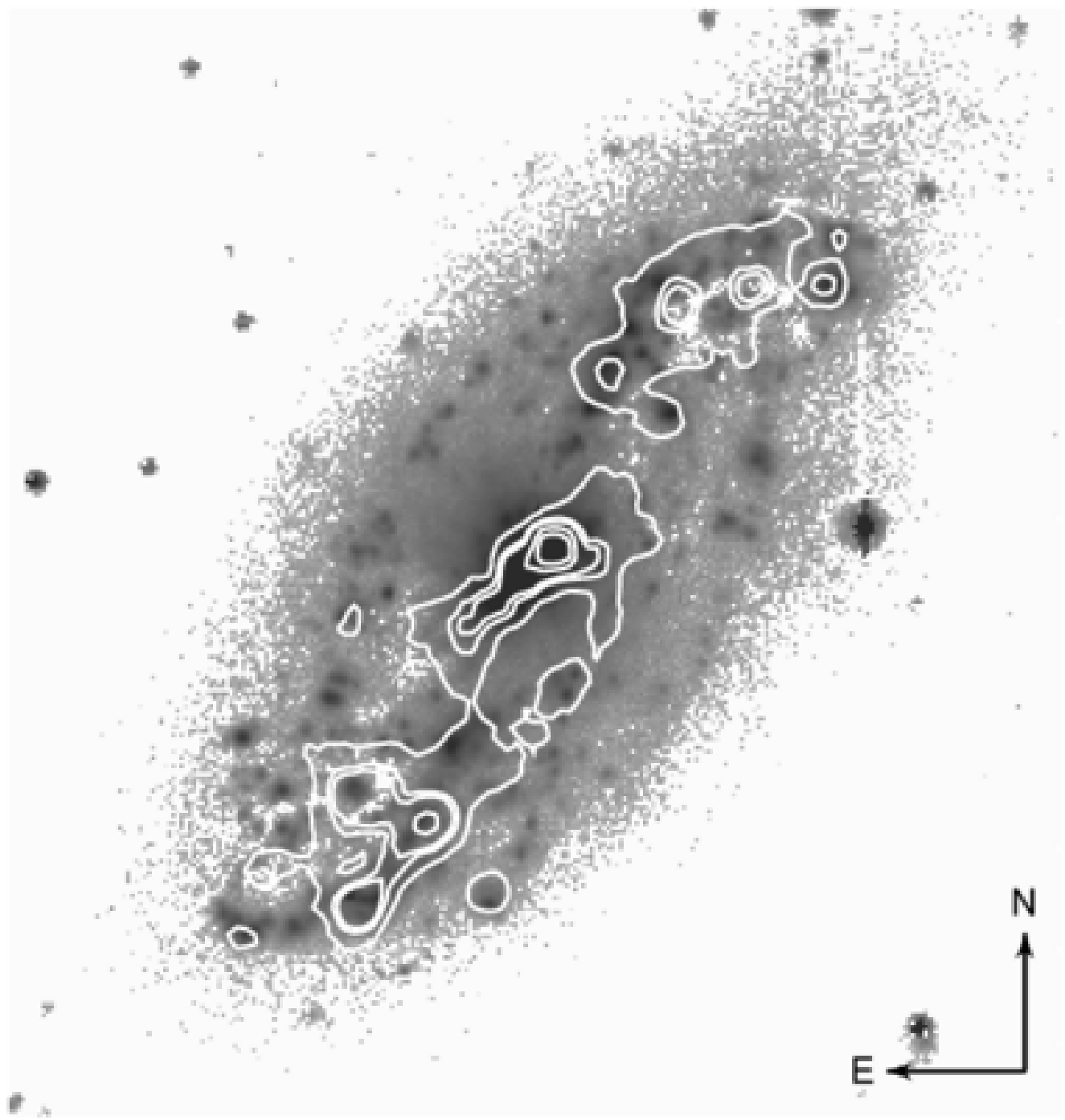,width=9cm,angle=0}}
\caption{Overlay of our UVW2 (212 nm) filter {\it OM} contours on top
the   EPIC  0.5--2.0   keV  frame   (left)  and    an H$\alpha$  image
(right).  (Contour   levels are  2,   7,  9, 10,   15,   20, 30, and  80
counts.) All images are displayed in logarithmic
scale with a threshold of 3$\sigma$ of the mean background value.}\label{fig:xuv}
\end{figure*}

\begin{figure*}[htb]
\centerline{
\psfig{file=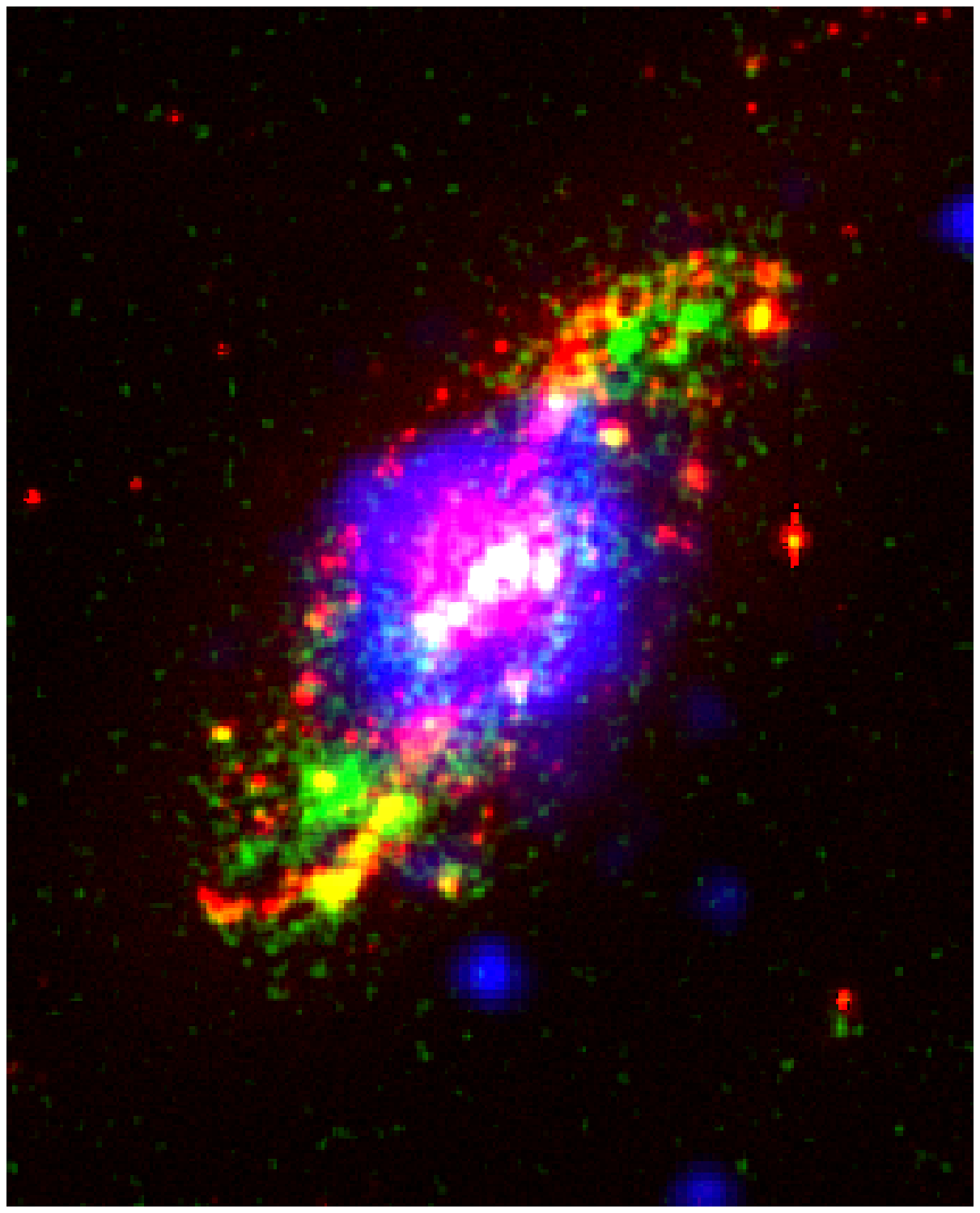,width=15cm,angle=0}}
\caption{RGB combination of \halpha~image (in red), UVW2 (in green),
and \epic\ (in blue). North is up and East is left. The dimensions of
the image are 5\arcmin$\times$3\farcm5.
}\label{fig:composite}
\end{figure*}

In Fig.~\ref{fig:xuv} we have overlaid the {\it OM} 212 nm image
of NGC\,1808 on the EPIC 0.5--2.0 keV frame from 
Fig.~\ref{fig:xbands} and on an extremely sensitive
H$\alpha$ image. An RGB composition of the
images in optical (\halpha), UV (\om\ UVW2) and X-rays (\epic) is
presented in Fig.~\ref{fig:composite}.

The correspondence of UV and H$\alpha$ emission maxima
indicates that the {\it OM} data trace the massive stars ionizing
the gas in \hii~regions in NGC\,1808. The H$\alpha$ image also shows several knots located in an outer spiral arm.
SF is traced  in the circumnuclear starburst but  also further out in
the galaxy disk.  Similarly, the soft  X-ray emission observed in  the
same regions  is expected to  come from  hot  gas in or  near these SF
regions. Optical/UV emission also tracks the bar.

\section{Discussion}

\asca\ and \ginga\ observations in the past missed the starburst's
soft  X-ray emission, favouring  an interpretation in  terms of an AGN
(Awaki et   al. 1996, Awaki \&  Koyama   1993), as  suggested  first by
V\'eron-Cetty and V\'eron (1985).  Similarly, the IR emission observed
by {\it ISO}\ is dominated by this AGN (Siebenmorgen et al. 2001).

\rosat, on the other hand, detected the soft emission from the
starburst, being less sensitive to the hard emission of an AGN
(e.g. Junkes et al. 1995), thus suggesting the predominance of
the starburst.

Based on  the data presented above,  the most likely interpretation is
that NGC\,1808 hosts both, a circumnuclear starburst and an unresolved
nuclear    X-ray     source.        Both    in    the    imaging
(Figs.~\ref{fig:xbands},\ref{fig:xcomp}, and \ref{fig:chandra_s1_s2})
and spectral (Figs.~\ref{fig:nucleus}, \ref{fig:spec_chandra}, and
\ref{fig:rgs_pn}) domains, \xmm\ and \chandra\ detect the presence
of the starburst and the hard unresolved nuclear  source. In the following we
discuss the physical origin of the emission of both types of activity:
the starburst and the unresolved X-ray source.

\subsection{X-ray emission of the inner region}
\label{par:centre}

\subsubsection{Hard component}

{\bf Spectral characteristics}

The slope of the power law fitted to the 16\arcsec\ region,
$\Gamma=0.79^{+0.07}_{-0.06}$, is flatter in comparison to results
obtained for radio-quiet AGN and LINERs.  Works by Georgantopoulos et
al. (2002) and Terashima et al. (2002) find values for the power law
spectral indices ranging from 1.7 to 2.3 for these objects. However,
there are also objects with lower spectral indices
(e.g. PKS~2251$+$113, with $\Gamma=0.95\pm0.24$, Reeves \& Turner,
2001).  Analysis of the \chandra\ data reveals that this hard emission
is dominated by one of the two detected nuclear sources, S2 in
Fig.~\ref{fig:chandra_s1_s2}, which is harder.  In addition, source
S2 is absorbed by a column density of
N$_H=3.1^{+0.8}_{-0.7}\times10^{22}$~cm$^{-2}$, compatible with values
observed in Compton-thin Seyfert~2 galaxies.  This fact suggests that
in case one of the nuclear peaks observed with \chandra\ is associated
with an AGN, the most likely possibility is that it is source S2.
Interestingly, the best-fit model of the
\xmm\ 16\arcsec\ region spectrum does   not require  absorption
in excess of the Galactic value. However, this
\xmm\ region   includes several emitting sources that could blur
the presence of an absorbing component in the best fit model.\\

\noindent{\bf Luminosity}
	
The total luminosity of the central region with a radius of 16\arcsec=
850~pc                  of                  NGC\,1808               is
L$_{\rm{2-10\,keV}}=(1.61\pm0.06)\times10^{40}$~erg     s$^{-1}$.  The
luminosity  in the same energy   band associated with  the harder peak
with  a radius of  1\farcs2=65~pc observed with  \chandra, i.e. source
S2,       is L$_{\rm{2-10\,keV}}=0.92^{+0.08}_{-0.7}\times10^{40}$~erg
s$^{-1}$, which  is $\sim60\%$  of   the emission of  the \xmm\  inner
region.  These values are in the  range of  Low Luminosity AGN (LLAGN)
and   LINERS   (Terashima et   al.     2002,  Georgantopoulos  et  al.
2002). Moreover, in  both  cases the luminosity  associated  with the
power  law component is in  good agreement with  the typical values of
low luminosity Seyfert galaxies and  LINERs.  Terashima et al.  (2002)
found in a sample of 53 objects of these types observed by \asca\ that
the          luminosity  varies         within        the        range
$L_{2-10\,keV}=5\times10^{39-40}$~ergs$^{-1}$.   If   only  LINERS are
considered,     the        mean  luminosity        obtained         is
$L_{2-10\,keV}=7\times10^{39}$~ergs$^{-1}$  (Ho et al. 2001), which is
lower than the one derived for NGC\,1808.  In the  soft band, a sample
of  LINERS observed   with    \rosat\  (Komossa  et  al.  1999)   shows
luminosities in the range of $L_{0.24-2\,keV}=10^{38-41}$~ergs$^{-1}$,
in  agreement with luminosity measured for the NGC\,1808.\\

\noindent{\bf Variability}

We studied the variability of NGC\,1808. While no short-term
variation of flux was  detected during the \xmm\ observation, medium- and
long-term variability has been suggested in both soft and hard
bands. 

NGC\,1808 was observed by \xmm\ and \chandra\ with an interval of
eight months.  The measurements indicate the  soft X-ray flux remained
constant, taking the uncertainties into account. A decrease of  16\%
(at a 2.5$\sigma$ level) in the measured flux is  observed in the hard
X-ray band. In  particular, the power  law component of  the innermost
region  spectrum of NGC\,1808 is responsible  for this variability.

The long-  and medium-term flux  variations measured  over the years by
various satellites are  most probably  caused by
variability   of  the nuclear source.    In the soft   energy band, a significant decrease of a factor of five has  been observed when the  \rosat\ and \xmm\  (or
\chandra) luminosities are compared.  It should  be  noted    that  the  \rosat\  spectral fit    is
particularly uncertain because of  the high absorbing column  density,
which leaves a very narrow effective  bandpass for the X-ray emission,
from about 0.5 to 2.4 keV.

In  summary, mild medium-term  variability of the hard band luminosity
was measured between  the \xmm\ and \chandra\ observations. This
variation can  be attributed  to  an AGN  or  a ULX.  Furthermore, the
uncertainties introduced  by  comparing different satellites does  not
allow us  to   firmly  conclude  that other   changes   in   flux  are
intrinsically due to the nucleus of NGC\,1808.

\noindent{\bf Ultraluminous X-ray sources}

The study of the hard X-rays emanating from the inner region indicates 
that the nuclear X-ray emission itself could be due to a LLAGN or a
LINER.    The      observed      hard  band     X-ray     luminosity,
L$_{2-10\,keV}$=$(1.61\pm0.06)\times10^{40}$~erg~s$^{-1}$,   is   very
difficult to explain by standard stellar processes. Nevertheless,
non-nuclear point-like sources with  X-ray luminosities on the order
of $10^{38-41}$~erg~s$^{-1}$ have been detected in nearby galaxies. An
alternative origin to an  AGN nucleus of NGC\,1808  could be one of the
so-called ultraluminous X-ray sources (ULX).

The  two  central sources S1 and  S2 were  studied  in detail in
Sect.~\ref{sec:chandraspec}.   The  nuclear   location  from  2MASS  is
compatible with either of them  being the  nucleus. The luminosity  of
source S2  of L$_{2-10~keV}=9.2^{+0.8}_{-7}\times10^{39}$~erg~s$^{-1}$
is compatible with either a LLAGN or a ULX.  Strickland et al. (2001)
and Roberts  et al.  (2002)  show that the  analysis  of \chandra\ ULX
spectra favours  simple  power  laws  with indices  in the  range   of
1.8--2.9,    higher than  the value  obtained    in  our analysis of
0.79$^{+0.07}_{-0.06}$.   Interestingly, one   of  the  compact  radio
sources  reported by  Saika et al.   (1990) coincides  with the  X-ray
point-like source S1.

Although  more data analysis  is  necessary to reach any   firm
conclusion,  this  finding leaves   open  the possibility  that the
nucleus of NGC\,1808 could be a ULX.

\subsubsection{Soft component}

{\bf Emission distribution}

The soft X-ray emission of  the inner region  of NGC\,1808 is extended
(see e.g. Fig.~\ref{fig:xbands}).  The elongated  emission is detected 
at energies up to  4.5~keV.  Moreover,  the \chandra\ images clearly
show the simultaneous presence of a double emission peak and diffuse, 
or at least unresolved, X-ray emission around the two peaks.  The
extended emission  is likely to have been created by star  formation processes.
Our \xmm\ data permit a clear spectral distinction between the emission
associated with the starburst and the additional  nuclear sources in the
r=16\arcsec\ (850~pc)   region,    see    Fig.~\ref{fig:nucleus}.   The
starburst, associated  with  the {\it  mekal} component in  the spectral
fit,    dominates  the   soft   energy   band   (below  1  keV),   see
Tables~\ref{tab:models} and~\ref{tab:lumin}.
The starburst  also shows up    in the form of  radio  continuum and IR
knots  (Saikia et al.  1990, Kotilainen et  al. 1996), bright \halpha\
emission, and bright circumnuclear  UV continuum from massive stars, as
displayed in Fig.~\ref{fig:xuv}. In particular,   the position of  the
2MASS nucleus   coincides  perfectly with one   of   the radio compact
sources detected by Saikia et al~(1990), explained as SNR or complexes
of unresolved  SNR.  The nucleus is  prominent  in X-rays  and  in the
optical regime (see Figs.~\ref{fig:xbands}\ and \ref{fig:uv}.a).

However,  while  the \om\  UVW2  image  shows  that the  UV
emission close   to the position   of the  nucleus is  very  weak (see
Fig.~\ref{fig:uv}.c).  These facts suggest  that the nucleus is highly
absorbed. The spectral analysis of source S2 shows a high value of the
equivalent   hydrogen  column,  N$_H\sim10^{22}$~cm$^{-2}$,  which can
explain this absence of UV and soft X-ray emission associated with the
hard X-ray peak.\\

\noindent{\bf Plasma temperature}

The best fit to the EPIC data contains a thermal plasma component
with a temperature of kT = $0.580^{+0.019}_{-0.016}$ keV. This corresponds roughly to 
the ``medium'' component identified in the spectra of other
starburst galaxies (Dahlem et al. 1998; see 
also Weaver et al. 2000  and Dahlem et al. 2000). 
It is consistent with $0.5\pm0.2$ keV (Junkes et al. 1995), 
which was measured from \rosat\ data, based  only on about 
600 photons. 
It is interesting to note that the  surprisingly high
Hydrogen column density, on the order of $8\times10^{21}$ cm$^{-2}$, 
in the best-fitting model to the \rosat\ data is also corroborated 
by our results.\\

\noindent{\bf X-ray luminosity and SFR}
\label{sec:sb_lum}

The luminosity of the starburst component, which  we identify with the
thermal spectral components, is dominated by the  emission in the soft
energy   band.            The        observed  X-ray      luminosity,
$L_{0.2-4\,keV}\sim1.3\times10^{40}$ erg  s$^{-1}$,    which represents
$\sim$70\% of the emission associated with the starburst component of the whole
galaxy, leads  to   a far-infrared-to-X-ray  luminosity   ratio of
log($L_{\rm    X}/L_{\rm   FIR})$ = --3.8~\footnote{$L_{\rm FIR}  =  2\pi\ D^2\
FIR=9.3\times10^{43}$~erg s$^{-1}$(Dahlem  et al.  1992), where $FIR =
1.26\  (2.58\ f_{60}  + f_{100}$),  $f_{60}$ being the  IRAS 60 $\mu$m
flux density and $f_{100}$ the 100 $\mu$m flux density.}. This
value is consistent with those obtained by Heckman et al. (1990) for a
sample of six starburst galaxies, which are in the range between --3.7
and  --4.3, and also with  the results in  Mas-Hesse et al. (1995), who
obtained a mean value of --3.33 for a sample of starforming galaxies.

Ranalli, Comastri \& Setti (2003) deduced that the X-ray luminosity is
a tracer of SFR,  based on a sample of   nearby galaxies.  As   was
pointed out by Persic et al.  (2004b), the luminosity in the 2--10~keV
band is  not  a precise SFR   indicator, due to   the contamination by
low-mass X-ray binaries (LMXB).  These objects are bright in the X-ray
band over long   timescales,  t$_X\sim10^7$~yr, and therefore   do not
trace the instantaneous  SFR.  For this reason, we  use the soft X-ray
luminosity as a   tracer of  the  SRF,  with the  SFR-L$_{0.5-2\,keV}$
relation in Ranalli, Comastri \& Setti (2003):

\begin{equation}
SFR_X(M_{\odot}\,yr^{-1})=2.2\times10^{-40}L_{0.5-2\,keV}
\end{equation}

The L$_{0.5-2\,keV}$   associated  with  the  starburst in    the  r=16\arcsec
(850~pc) region,
as derived  from the mekal  component  contribution, is on the
order of  L$^{SB}_{0.5-2\,keV}\sim1\times10^{40}$ erg~s$^{-1}$ and the
SFR$_X$,   therefore, is $\sim2.5$~\msolyr. \\

\noindent{\bf Metallicities}

We detect several key emission lines, which for the first time provide
us with    a  good  measure   of  the abundances    of   some elements
(Sect.~\ref{par:epicspec}).    The \rgs\ spectrum  shows   a very   weak
continuum emission with several  emission lines.  The values  obtained
indicate  slightly  super-solar metallicities for   Ne, Mg, and Si and
solar or  sub-solar  abundance for Fe,  which is  consistent with the
results by Dahlem et al. (1998) and Weaver at al (2000).

In  this context  the  absence  of  a  thermal  Fe  K$\alpha$ line  at
$\sim$6.5 keV  energy is  noteworthy.   From our  data in the  annular
region, we derive a 3-$\sigma$ upper limit on the equivalent width of
170~eV  and on the flux of $4\times10^{-7}$~photons~cm$^{-2}$s$^{-1}$.
This  result  is compatible with any   possible iron line emission 
entirely due to  the mekal component. We  have used  ATOMDB~v.1.3 to
derive  the  expected  flux  of  the emission   lines  for a  gas with
temperature and normalisation as  obtained from the single mekal
component     fitted     to     the     annular    spectrum,      i.e.
kT=$0.53^{+0.05}_{-0.03}$~keV              and            normalisation     A$_{\rm
mekal}=2.2\times10^{-4}$. The total flux  of  the most prominent  iron
lines    (FeXXI-6.505~keV   and  FeXXII-6.504   \&   6.57~keV  )  is
$\sim2\times10^{-4}$~photons~cm$^{-2}$s$^{-1}$ (with an emissivity  of
$\sim5\times10^{-22}$~photons~cm$^{3}$s$^{-1}$), consistent  with  the
upper limit measured.

\subsection{X-ray emission outside the central region}
\label{par:outer}

In Sect.~\ref{sec:spec_ana}, we show the results of the spectral
analysis of the 875~pc--1.9~kpc annular region. The best fit model 
indicates that, apart from the contribution of the nuclear emission, 
the annular region includes an extra thermal component which can be 
explained by a mekal model with kT=$0.10^{+0.02}_{-0.01}$~keV. 
The analysis of the total galaxy X-ray emission also indicates the 
presence of a soft thermal model with a similar temperature,
kT=$0.11^{+0.03}_{-0.02}$~keV.

Although weak, the contribution   in the soft  band of  this component
(L$_{0.1-2.4\,keV}=7.7\times10^{39}$~erg~s$^{-1}$) to the  soft X-ray
luminosity  of the whole   galaxy is significant, representing  around
30\% of  the total soft  X-ray  emission of NGC\,1808.  In  the  hard
band,  the contribution of this  low  temperature thermal component is
negligible.  Using Eq.\,1
(see Sect.~\ref{sec:sb_lum} for  further details), the observed X-ray
luminosity  associated with both  mekal components  of the annular
region,  L$_{0.5-2\,keV}\sim4\times10^{39}$,  can be translated into a
SFR of $\sim$1~\msolyr.

Far-infrared emission is also a good tracer of the star formation; using
the relationship by Kennicutt et al. (1998), we derive for NGC\,1808:

\begin{equation}
SFR_{FIR}(M_{\odot}\,yr^{-1})=\frac{1}{e}4.5\times10^{-44}L_{FIR}
\end{equation}

\noindent where $e$ is the fraction of UV/optical flux  emitted by the stars and
absorbed by the dust and  remitted in the IR,  which is close  to
unity   even   for  low   values of   reddening   (Mas-Hesse \& Kunth
1991). Assuming  that the whole FIR  emission is due to the starburst,
we obtain  SFR$_{FIR}\sim$4~\msolyr. This  result  is consistent with
the one   obtained through the X-ray  emission,  combining the values
derived   from  the  nuclear   region  (SFR$_{X}\sim2.5$~\msolyr,  see
Sect.~\ref{sec:sb_lum})  and      from     the     annular      region
(SFR$_{X}\sim1$~\msolyr). It suggests that the contribution of the AGN 
to the FIR emission is small.

The second, soft mekal component detected in both the annular and 
the r=3.3~kpc region is most likely thermal emission associated with the
halo. Such a thermal   component  (kT=0.2--0.3~keV) is  found in   all
starburst galaxies in the  sample by Dahlem et  al.  (1998) and Weaver
et al. (2000) and also by us in other starburst galaxies observed with
\xmm\  (e.g. Dahlem   et  al.  2003 and    Ehle \& Dahlem 2004).   The
presence of both, the  prominent vertical dust filaments  in NGC\,1808
and   the  gaseous outflow  inferred by  the   kinematics  of \hi\ gas
(Koribalski et  al. 1993),  suggested that  this component  would most
likely be present in NGC\,1808.

If the diffuse X-ray emission in the soft energy band is assumed to be
due to   thermal cooling  of  a   hot gas   in collisional  ionization
equilibrium\footnote{See, however, the   discussion in the validity of   this
assupntion in Breitschwerdt  \& Schmutzler (1999)},  it is possible to
estimate the gas density  $n_{\rm e}$, mass  $m_{\rm gas}$, and cooling
time $\tau$ of  that  plasma.  To this end   we make use of  a formula
given by Nulsen   et al.  (1984),  i.e.   $L_{\rm x}({\rm soft})=0.812
\Lambda(T)  n_{\rm e}^2  V \eta$.   The  unknown filling factor $\eta$
allows for some clumpiness of the gas filling the emitting volume V,
and was found in simulations (Avillez
\& Breitschwerdt, 2004  and references therein) to be  in the range of
0.2--0.4, only modestly increasing with higher SN rate.

As explained in Sect. 3.2.2,  the spectrum of the non-nuclear extended
X-ray emission  is complex and a  best fit was obtained  with a
model consisting of a foreground  absorbed power law plus (at least) two
internally absorbed mekal components,  corroborating the fact that the
hot gas is a  multi-temperature mixture. Nevertheless for the  purpose
of estimating  hot gas parameters,  we  fitted this emission component
with  a  single  'characteristic' temperature that   was  found  to be
$0.53^{+0.05}_{-0.03}$~keV.     For such a  gas    temperature of  the
starburst component ($\sim6.2\times10^6$~K) Raymond et  al.  (1976) give a
cooling coefficient  $\Lambda(T)$   of  $\sim5\times10^{-23}$   erg  cm$^3$
s$^{-1}$. For the unabsorbed luminosity of the hot gas component we
make   use       of   the   total   ({\it       Thermal    I  +   II})
$L_{0.1-2.4~keV}=7.7\times10^{39}$~erg~s$^{-1}$                   (see
Table~\ref{tab:lumin}).   Assuming for the   emitting volume  a  torus
geometry with  an inner radius of 875~pc  and outer radius of 1.9~kpc,
the calculated gas parameters are presented in Table~\ref{gas}.

\begin{table}
\caption{Parameters for the hot gas component in the extra-nuclear starburst region of NGC\,1808}
\label{gas}
\begin{flushleft}
\begin{tabular}{ccc}
\hline
\noalign{\smallskip}
$n_{\rm e}$&$m_{\rm gas}$&$\tau$\\
(cm$^{-3}$)&$(M_{\sun})$&(yr)\\
\noalign{\smallskip}
\hline
\noalign{\smallskip}
$3.0\times10^{-2}/\sqrt{\eta}$ & $6.3\times10^6 \sqrt{\eta}$ & $6.3\times10^7 \sqrt{\eta}$\\
\noalign{\smallskip}
\hline
\end{tabular}
\end{flushleft}
\end{table}

The mass of  the detected hot  gas is only  $\la0.5\%$  of the  total gas
mass\footnote{The total  gas mass calculated using  H$_I$ observations
is   $>1.8\times10^9$\msol (Dahlem et al. 2001).}.  The  cooling  time,
$\tau$, is   comparable to typical duty cycles of starbursts, which are 
estimated to be on the order of a few times $\sim10^7$ yr (Rieke et al. 1988).

No significant  X-ray emission  was  found, at the sensitivity  of our
current data, in the intergalactic space around NGC\,1808.  Similarly,
no \hi\ and \halpha\ emission was found by us (Dahlem et al. 2001).

\section{Summary and conclusions}
\label{par:summ}

The  work presented in this paper  shows evidence of the presence
of starburst activity and a hard  unresolved source within the inner
r=850~pc region in NGC\,1808.  Analysis of the optical-to-X-ray
emission proves the co-existence  of   thermal diffuse plasma,    non
nuclear point-like sources and, of a LLAGN or a ULX.

The \xmm\ EPIC images show extended X-ray  emission for energies below
4.5~keV. The  elongated shape observed  in the  X-ray band follows the
same orientation  as the   H$\alpha$  and optical-UV    emission,
suggesting that  SF  contributes significantly    to the soft   X-ray
emission.  At least  four non-nuclear point-like sources  were
detected in the \chandra\ ACIS image of the inner 250~pc of
NGC\,1808. The high spatial resolution \chandra\ image allows the
detection of a double-peak nuclear structure. The locations of both
sources are compatible with the position of the nucleus as derived
from 2MASS. The spectral characteristics of these sources show that one
is dominated by hot gas emission, indicating starburst activity, while
the other is a highly absorbed hard X-ray source.

The spectral analysis of the \xmm\ data  completes the scenario outlined 
by X-ray imaging. Comparing  \pn\ spectra of several regions
in the central part of NGC\,1808, it is  inferred that the bulk of the
emission originates from the nucleus,  although in the soft X-ray band
the contribution of the circumnuclear regions is  not negligible.

The  X-ray \epic\ spectrum of the  inner nuclear region (r=16\arcsec, 
850 pc) is explained
by a power law, $\Gamma=0.79^{+0.07}_{-0.06}$, accounting for the hard
X-ray emission   and  associated  with   an unresolved    point-like
source. The absorption measured in the \chandra\ spectral analysis
of   source S2 is  compatible with typical    values  of Compton-thin
Seyfert~2 galaxies. However,  this absorbing material does not show up
in the \xmm\ spectrum, probably because it includes contributions from
different regions, making it difficult to separate absorption and soft
emission components.

In order to explain the  soft X-ray emission of
the inner region, it is necessary to  include emission due to a diffuse
plasma  with kT=0.580$^{+0.019}_{-0.016}$~keV.  The data obtained with
\rgs\ exhibit   emission  lines   similar in  wavelength   and relative
intensity ratios to   the  ones found for  the  prototypical starburst
galaxy M\,82.  The temperature, the  luminosity, and the various ratios
of emission lines are consistent  with being originated in a  collisionally 
ionized plasma associated with SF regions, which dominates the total 
emission spectrum including the continuum in the soft X-ray regime.

The luminosities measured with \xmm\ for the r=850~pc inner region 
(L$_{2-10\,keV}=(1.61\pm0.06)\times10^{40}$~erg~s$^{-1}$) and with
\chandra\ for source S2
(L$_{2-10\,keV}=0.92^{+0.08}_{-0.7}\times10^{40}$~erg~s$^{-1}$) are
both in the range of LLAGNs but are also compatible with the values
of ULX sources. A decrease in the 2-10~keV flux on the order of  15$\%$ 
in an interval of eight months was measured when comparing \xmm\ 
and \chandra\ observations, while no change in flux was detected in 
the soft energy band. Both the luminosities  and the variability
detected are compatible with the unresolved source being a LLAGN or 
a ULX.

Halo  emission of  NGC\,1808  could also be detected: a  softer thermal
component,  kT$\sim$0.1~keV,  with  L$_{0.1-2.4\,keV}\sim7\times10^{39}$~erg~s$^{-1}$
($\sim$30\% of the total soft X-ray  emission) is necessary to explain
the spectrum at distances larger than 875 pc from the centre.

The   multiwavelength analysis  of the  nuclear   region  of NGC\,1808
performed  in this work   allows us to detect  and for the first time  to
unambiguously disentangle the  contributions of an unresolved  nuclear
X-ray source and  the starburst regions   within the r=850~pc  region.
Further  analysis of better quality data  is required to determine the
exact nature of the unresolved nuclear source.

\acknowledgements{ We specially thank N. Stuhrmann for
reducing the H$\alpha$ data. EJB, MSLl and JMMH have been supported by
Spanish MEC under grant AYA2001-3939-C03-02}

\end{document}